\documentclass[journal,twoside,a4]{IEEEtran}

\IEEEoverridecommandlockouts

\usepackage{cite}
\usepackage{amsmath,amssymb,amsfonts}
\usepackage{mathtools} 
\usepackage{graphicx}
\usepackage{textcomp}
\usepackage{xcolor}
\usepackage{tabularx}
\usepackage{booktabs}
\usepackage{float}
\usepackage[english]{babel}
\usepackage{graphicx,lipsum}
\usepackage{paralist}
\usepackage{array}
\usepackage{amsmath,stackrel}
\usepackage{multicol}
\usepackage{verbatim}
\usepackage{enumerate}              
\usepackage{dsfont}
\usepackage{psfrag}
\usepackage{algorithm}
\usepackage[nolist]{acronym}
\usepackage[utf8]{inputenc}
\usepackage{tikz}
\usepackage{collcell}
\usepackage{pgfplots}
\pgfplotsset{compat=1.18}
\usepackage{hhline}
\usepackage{colortbl}
\usepackage{multido}
\usepackage{multirow}
\usepackage{dblfloatfix}    
\usepackage{etoolbox}
\usepackage{soul}
\usepackage{eucal,balance}

\begin{acronym}
\acro{AWGN}{additive white Gaussian noise}
\acro{AoA}{angle of arrival}
\acro{AoD}{angle of departure}
\acro{BF}{beamformer}
\acro{BS}{base station}
\acro{CIR}{channel impulse response}
\acro{CP}{cyclic prefix}
\acro{CKF}{cubature Kalman filter}
\acro{CPHD}{cardinalized probability hypothesis density}
\acro{DBSCAN}{density-based spatial clustering of applications with noise}
\acro{DoA}{direction of arrival}
\acro{DoD}{direction of departure}
\acro{EIRP}{effective isotropic radiated power}
\acro{ELP}{equivalent low-pass}
\acro{ESPRIT}{Estimation of Signal Parameters via Rotational Invariance Techniques}
\acro{FC}{fusion center}
\acro{FD}{frequency division}
\acro{FFT}{fast Fourier transform}
\acro{FDD}{frequency-division duplexing}
\acro{GM}{Gaussian mixture}
\acro{GMPHD}{Gaussian mixture probability hypothesis density}
\acro{GMCPHD}{Gaussian mixture cardinalized probability hypothesis density}
\acro{GOSPA}{generalized optimal subpattern assignment}
\acro{ICI}{inter-carrier interference}
\acro{InF}{indoor factory}
\acro{IoT}{internet of things}
\acro{IFFT}{inverse fast Fourier transform}
\acro{ISI}{inter-symbol interference}
\acro{i.i.d.}{independent, identically distributed}
\acro{JSC}{joint sensing and communication}
\acro{k-NN}{k-nearest neighbors}
\acro{LOS}{line-of-sight}
\acro{LTE}{long term evolution}
\acro{MAP}{maximum a posteriori}
\acro{MB}{multi-Bernoulli}
\acro{MBM}{multi-Bernoulli mixture}
\acro{MCL}{maximum coupling loss}
\acro{MPL}{maximum path loss}
\acro{MIL}{maximum isotropic loss}
\acro{MC}{Monte Carlo}
\acro{MIMO}{multiple-input multiple-output}
\acro{MUSIC}{MUltiple SIgnal Classification}
\acro{MDL}{minimum description length}
\acro{MHT}{multi-hypothesis tracker}
\acro{mmWave}{millimeter-wave}
\acro{NR}{new radio}
\acro{NLoS}{non-line-of-sight}
\acro{OFDM}{orthogonal frequency division multiplexing}
\acro{OSPA}{optimal sub-pattern assignment}
\acro{p.d.f.}{probability density function}
\acro{PF}{particle filter}
\acro{PHD}{probability hypothesis density}
\acro{PSD}{power spectral density}
\acro{PPP}{Poisson point process}
\acro{QPSK}{quadrature phase shift keying}
\acro{QuaDRiGa}{QUAsi Deterministic RadIo channel GenerAtor}
\acro{RCS}{radar cross-section}
\acro{RF}{radio frequency}
\acro{RFS}{random finite set}
\acro{RMSE}{root mean squared error}
\acro{r.v.}{random variable}
\acro{RRH}{remote radio head}
\acro{SD}{space division}
\acro{SSIR}{signal-to-self interference ratio}
\acro{SI}{self interference}
\acro{SNR}{signal-to-noise ratio}
\acro{SU}{single-user}
\acro{SCM}{sample covariance matrix}
\acro{SPF}{soft particle filter}
\acro{TD}{time-division}
\acro{TDD}{time-division duplexing}
\acro{TDOA}{time difference of arrival}
\acro{UMi}{urban micro}
\acro{UE}{user equipment}
\acro{ULA}{uniform linear array}
\acro{V2V}{vehicle-to-vehicle}
\end{acronym}

\DeclarePairedDelimiter{\nint}\lfloor\rceil
\DeclareMathOperator*{\argmax}{argmax}

\def\undertilde#1{\mathord{\vtop{\ialign{##\crcr
$\hfil\displaystyle{#1}\hfil$\crcr\noalign{\kern1.5pt\nointerlineskip}
$\hfil\tilde{}\hfil$\crcr\noalign{\kern1.5pt}}}}}

\definecolor{MATLAB_blue}{rgb}{0, 0.4470, 0.7410}
\definecolor{MATLAB_yellow}{rgb}{0.9290, 0.6940, 0.1250}
\definecolor{MATLAB_red}{rgb}{0.8500, 0.3250, 0.0980}
\definecolor{MATLAB_purple}{rgb}{0.4940, 0.1840, 0.5560}
\definecolor{MATLAB_green}{rgb}{0.4660, 0.6740, 0.1880}
\definecolor{MATLAB_azure}{rgb}{0.3010, 0.7450, 0.9330}

\begin{document}

\title{Sensor Fusion and Resource Management in MIMO-OFDM Joint Sensing and Communication}

\author{Elia~Favarelli,~\IEEEmembership{Member,~IEEE,}        Elisabetta~Matricardi,~\IEEEmembership{Student Member,~IEEE,}
Lorenzo~Pucci,~\IEEEmembership{Student Member,~IEEE,}
Wen~Xu,~\IEEEmembership{Senior Member,~IEEE}, 
Enrico~Paolini,~\IEEEmembership{Senior Member,~IEEE,}
and Andrea~Giorgetti,~\IEEEmembership{Senior Member,~IEEE}%
\thanks{This work was presented in part at the Euroepan Radar Conference (EuRAD), Berlin, Germany, Sept. 2023.\\
\indent This work was supported in part by the CNIT National Laboratory WiLab and the WiLab-Huawei Joint Innovation Center and in part by the European Union under the Italian National Recovery and Resilience Plan (NRRP) of NextGenerationEU, partnership on ``Telecommunications of the Future'' (PE00000001 - program ``RESTART'').\\
\indent E. Favarelli, E. Matricardi, L. Pucci, E. Paolini and A. Giorgetti are with the Dept. of Electrical, Electronic, and Information Eng. ``Guglielmo Marconi'' (DEI), University of Bologna, and WiLab, CNIT, Italy (e-mail: \{elia.favarelli, elisabett.matricard3, 
 lorenzo.pucci3, e.paolini, andrea.giorgetti\}@unibo.it).\\
W. Xu is with the Munich Research Center, Huawei Technologies Duesseldorf GmbH, Germany (e-mail: wen.dr.xu@huawei.com).}
}

\markboth{}%
{Favarelli \MakeLowercase{\textit{et al.}}: Sensor Fusion and Resource Management in MIMO-OFDM Joint Sensing and Communication}

\maketitle

\begin{abstract}
This study explores the promising potential of integrating sensing capabilities into \ac{MIMO}-\ac{OFDM}-based networks through innovative multi-sensor fusion techniques, tracking algorithms, and resource management. A novel data fusion technique is proposed within the \ac{MIMO}-\ac{OFDM} system, which promotes cooperative sensing among monostatic \ac{JSC} base stations by sharing range-angle maps with a central fusion center. To manage data sharing and control network overhead introduced by cooperation, an excision filter is introduced at each base station. After data fusion, the framework employs a three-step clustering procedure combined with a tracking algorithm to effectively handle point-like and extended targets.
Delving into the sensing/communication trade-off, resources such as transmit power, frequency, and time are varied, providing valuable insights into their impact on the overall system performance. Additionally, a sophisticated channel model is proposed, accounting for complex urban propagation scenarios and addressing multipath effects and multiple reflection points for extended targets like vehicles.
Evaluation metrics, including \ac{OSPA}, downlink sum rate, and bit rate, offer a comprehensive assessment of the system's localization and communication capabilities, as well as network overhead.
\end{abstract}


\IEEEpeerreviewmaketitle

\section{Introduction}
\IEEEPARstart{T}{he} prominence of safety in urban environments has recently emerged as a critical area of interest. This requires the integration of advanced sensing technologies. The growing interest in sensing stems from its potential to enable several applications, including but not limited to, traffic monitoring, autonomous driving, and environmental mapping. These applications are poised to benefit greatly from effectively utilizing radio frequency signals for sensing purposes \cite{BarLiuWinCon:J22}. 
Of particular importance is the advancement in sensing capabilities facilitated by the transition to higher frequency bands and the use of larger antenna arrays \cite{Tho:C21, Sch:C20}. This trend seems to perfectly capture the vision of the next generation of mobile networks, which are expected to encompass a wide range of novel capabilities and services, among which sensing seems one of the most prominent.

The sharp increase in the number of \acp{BS} opens the door to the development of precise and reliable localization systems \cite{KwoLiuConParWin:J23,MorRazWinCon:J23}. This, in turn, paves the way for enhanced safety measures in urban settings \cite{FenWeiChe:21}. Additionally, the advent of massive \ac{MIMO} technology at \ac{mmWave} frequencies offers the opportunity not only to achieve remarkable data transmission rates but also to attain accurate object position estimation \cite{JohVenGroLop:J22}. This noteworthy capability is anticipated to address emerging communication challenges effectively, embracing crucial aspects such as precise beam management \cite{Cui:21IntegratingSA}, accurate estimation of target direction for beam tracking to maintain high-quality links \cite{LiuZhaoWang:19}, and the application of predictive beam tracking using \ac{JSC} in vehicular scenarios \cite{ZhenLiu:22}.

Moreover, as we delve into the potential of 6G mobile radio networks and their applications, it is essential to examine the existing gaps in research. While significant strides have been made in integrating sensing and communication capabilities, certain aspects remain underexplored. In particular, the exploitation of mobile radio networks as an enabler for distributed and cooperative sensing is still underinvestigated. Along these lines, the benefits of cooperation, as well as its impact on network overhead, have not been thoroughly addressed. Furthermore, the remarkable flexibility offered by mobile networks to allocate resources based on user needs has not been fully exploited for sensing, hindering adjustment of the sensing/communication trade-off to specific requirements \cite{Favarelli:C23,Favarelli:D23,Favarelli:E23}. The integration of cooperative sensing into the mobile network framework stands as an unexplored frontier, promising to enhance both the efficiency of sensing applications and the overall performance of the network.

\begin{figure}[t]
    \centering    
    \includegraphics[width=1\columnwidth]{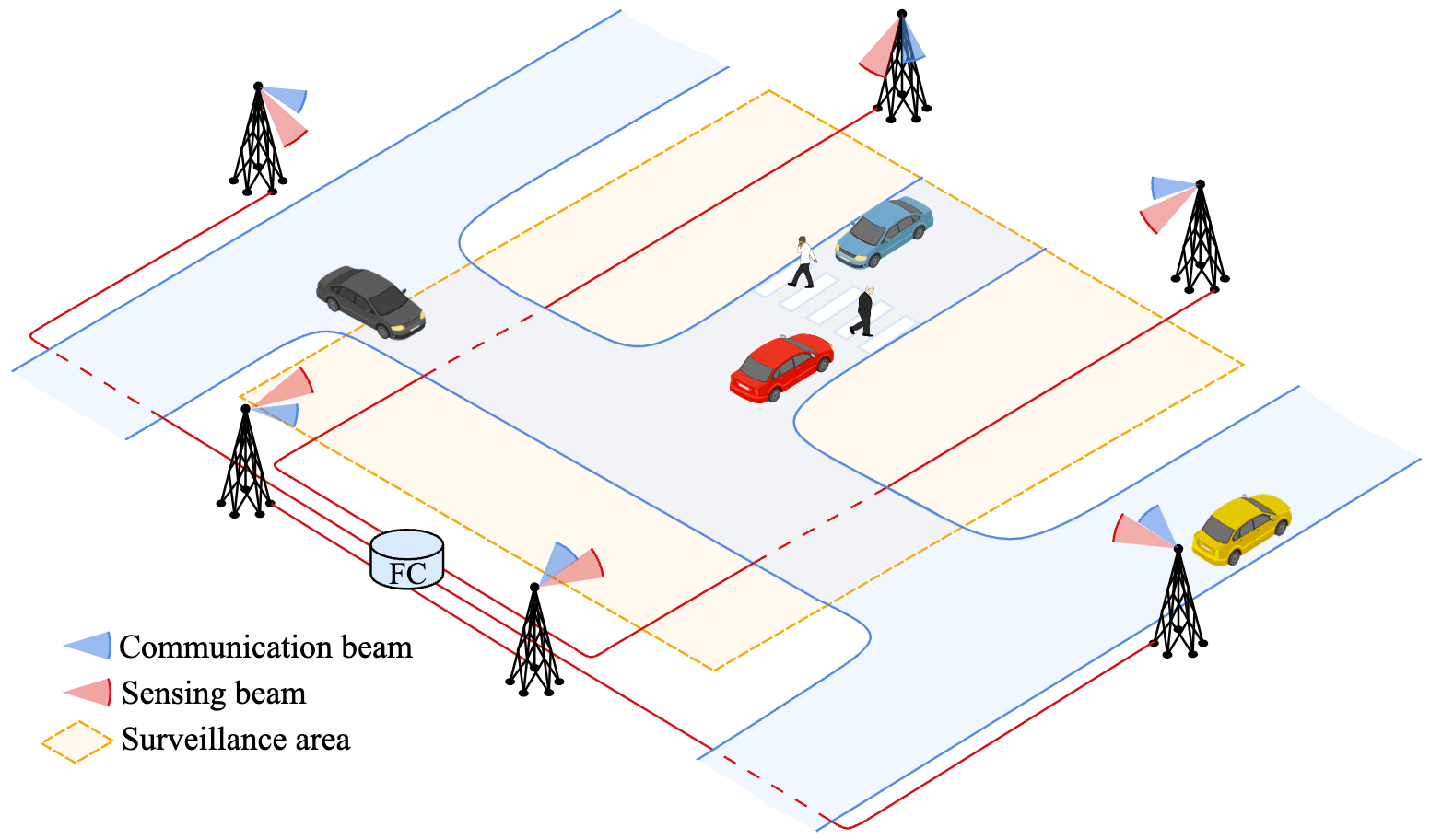}
    \caption{An urban scenario with six monostatic \ac{JSC} \acp{BS} aiming at monitoring pedestrians (point-like targets) and vehicles (extended targets) in a surveillance area. \acp{BS} communicate with their \acp{UE} while simultaneously sensing the surrounding environment via dedicated sensing beams. The \ac{FC} collects measurements from the \acp{BS} via the backhaul network, fuses them to create likelihood maps, and performs detection and multiple target tracking.}
    \label{fig:scenario}
\end{figure}

This study explores the feasibility of enabling sensing in \ac{MIMO}-\ac{OFDM}-based networks through the utilization of multi-sensor fusion techniques, tracking algorithms, and resource management. Specifically, we aim at exploiting range-angle radar maps derived from a collaborative set of \acp{BS} to track heterogeneous targets in a surveillance area. The main contributions of this work are as follows:

\begin{itemize}
\item With reference to a \ac{MIMO}-\ac{OFDM} system, we propose a data fusion technique to facilitate cooperative sensing among monostatic \ac{JSC} \acp{BS} by exchanging soft maps with a \ac{FC}.
\item We propose a thresholding strategy to control the amount of data shared between \acp{BS} and \ac{FC}.
\item We design a tailor-made three-step clustering procedure followed by a tracking algorithm to manage extended targets; we compare two tracking strategies, a \ac{PHD} filter and a \ac{MBM} filter.
\item We explore the sensing/communication trade-off at the \ac{BS} by varying the fraction of resources—power, frequency, and time—and provide insight into their impact on sensing and communication performance.
\item To account for the complex channel propagation in an urban environment, we propose a channel model that considers multipath propagation in the \ac{BS}-target-\ac{BS} path and multiple reflection points to address extended targets such as vehicles.
\item We assess the overall system capabilities using \ac{OSPA}, downlink sum rate, and bit rate to evaluate localization performance, communication capabilities, and network communication overhead, respectively.
\end{itemize}

Capital and lowercase boldface letters denote matrices and vectors, respectively; $\mathbf{I}_n$ is the $n\times n$ identity matrix; $\| \cdot \|_p$ stands for the $p$-norm; $\nint{\cdot}$ represents the round operator; $\delta(\cdot)$ is the Dirac delta function; $(\cdot)^*$ stands for the conjugate, and $(\cdot)^\mathsf{T}$ represents the transpose operation. A zero-mean circularly symmetric complex Gaussian random vector with covariance $\boldsymbol{\Sigma}$ is denoted by $\mathbf{x} \thicksim \mathcal{CN}( \mathbf{0},\boldsymbol{\Sigma})$, and $\mathbf{x} \thicksim \mathcal{N} (\boldsymbol{\mu},\boldsymbol{\Sigma})$ denotes the real-valued Gaussian random vector with mean $\boldsymbol{\mu}$ and covariance $\boldsymbol{\Sigma}$.

The rest of the paper is organized as follows. Section~\ref{sec:system} presents the \ac{JSC} network for cooperative sensing system, focusing on the description of the system, channel, and target models. Section~\ref{sec:tracking} describes the data fusion strategy, the clustering scheme, and the tracking algorithms. The performance of the proposed scheme is evaluated in Section~\ref{sec:results}, and conclusions are drawn in Section~\ref{sec:conclusion}.

\section{System Model}\label{sec:system}
We consider a \ac{JSC} network, illustrated in Fig.~\ref{fig:scenario}, which consists of multiple  \acp{BS} operating at \ac{mmWave} frequencies with monostatic sensing capability. Each \ac{BS} transmits an \ac{OFDM} waveform via multiple beams. In particular, while performing environment sensing through a dedicated beam and downlink signals, each \ac{BS} simultaneously communicates with its \acp{UE} present in the scenario. Through the backhaul network, the \acp{BS} cooperate with a \ac{FC} to accomplish targets' detection and tracking via data fusion. To mitigate inter-cell interference, \acl{FD} and \acl{TD} are exploited among the \acp{BS} so that sensing beams do not interfere; this operation is usually performed also for communication.

To sense the surrounding environment, each \ac{BS} uses a fraction of the available resources in the downlink. More precisely, the \ac{BS} transmits multiple frames each consisting of $M_\mathrm{s}$ \ac{OFDM} symbols with
\begin{equation}\label{eq:rho_f}
    \rho_\mathrm{f}=K_\mathrm{s}/K_0
\end{equation}
the fraction of subcarriers for \ac{JSC} among the total $K_0$ available. The \ac{OFDM} symbol duration $T_\mathrm{s}$ and the subcarrier spacing $\Delta f$ are kept fixed.
%
Multiple scans (measurements) of the surveillance area are performed to track the trajectory of the targets, with the time between two consecutive scans, $T_\mathrm{meas}$, that can be varied to accommodate the communication/sensing trade-off. Hence, we indicate with
\begin{equation}\label{eq:rho_t}
    \rho_\mathrm{t}=T_\mathrm{scan}/T_\mathrm{meas}
\end{equation}
the fraction of time reserved for \ac{JSC}. 

\begin{figure*}
    \centering
    \includegraphics[width=1\textwidth]{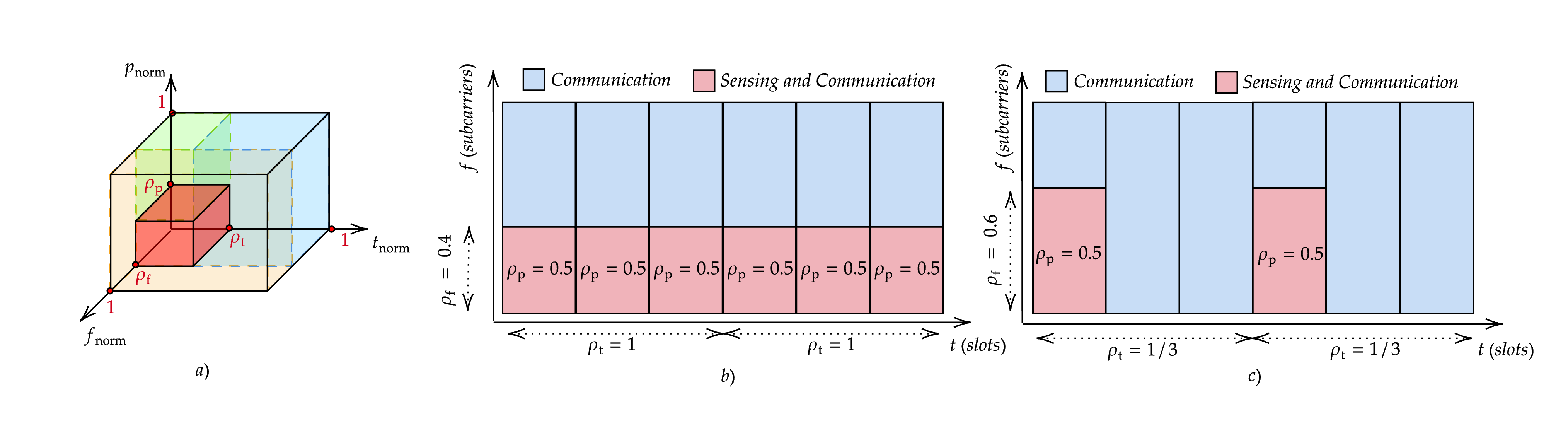}
    \caption{a) Graphical representation of the resource partition between communication and sensing at the \ac{BS}. The red block represents the fraction of resources reserved for sensing, and the other blocks represent the remaining resources available for communication. b) Example of sensing and communication resource allocation in the time-frequency domain, with $\rho_\mathrm{p}=0.5$, $\rho_\mathrm{f}=0.4$, and $\rho_\mathrm{t}=1$. c) Another example with $\rho_\mathrm{p}=0.5$, $\rho_\mathrm{f}=0.6$, and $\rho_\mathrm{t}=1/3$.}
\label{fig:block}
\end{figure*}

Consequently, the transmitted signal in \ac{JSC} slots 
can be represented by a matrix $\mathbf{X} \in \mathbb{C}^{K_\mathrm{s} \times M_\mathrm{s}}$, where the elements $x_{k,m}$ correspond to complex modulation symbols \cite{PucPaoGio:J22}.

\begin{figure*}
    \centering
    \includegraphics[width=2\columnwidth]{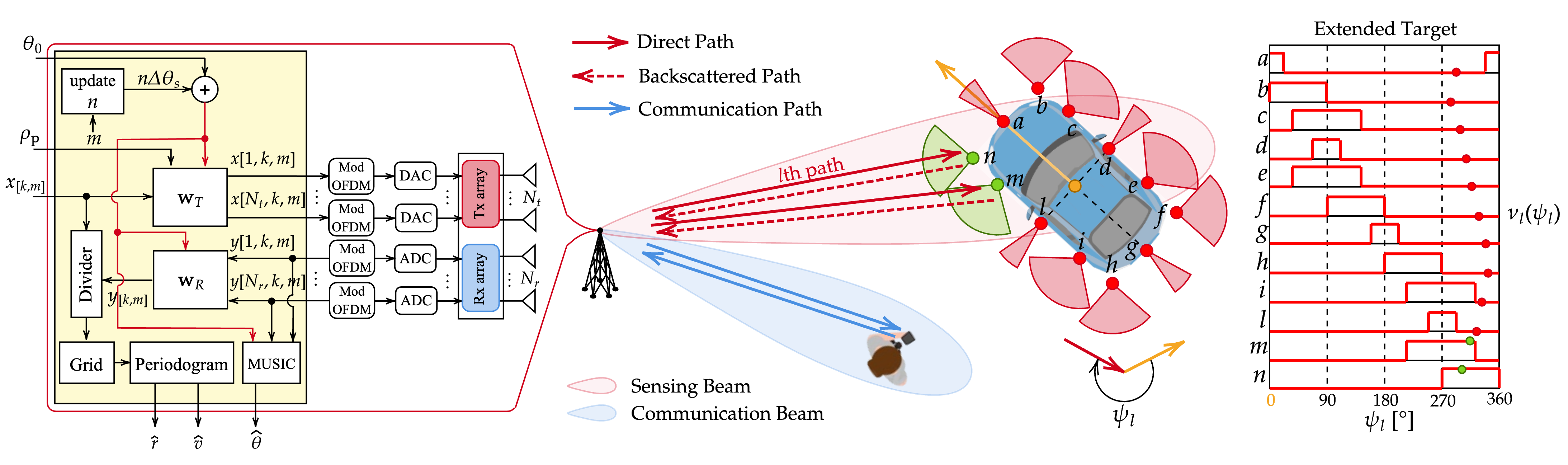}
    \caption{The \ac{MIMO}-\ac{OFDM} monostatic \ac{BS} with multi-beam capability and a typical target vehicle. The vehicle is represented as an extended target made of distributed scatterers with specific visibility functions represented on the right.}
    \label{fig:multipath}
\end{figure*}

Without loss of generality, each \ac{BS} is equipped with two \acp{ULA} with half wavelength spacing, one with $N_\mathrm{T}$ antennas for transmission and the other with $N_\mathrm{R}$ antennas for reception (see Fig.~\ref{fig:multipath}). 

At the transmitter, the beamforming vector $\mathbf{w}_\mathrm{T} \in \mathbb{C}^{N_\mathrm{T} \times 1}$ is applied to data symbols at each antenna, resulting in $\mathbf{x}[k,m] = \mathbf{w}_\mathrm{T} x_{k,m}$ \cite{asplund2020advanced}. Hence the communication beam (directed towards the \ac{UE}) and the sensing beam (steered to scan the area) \cite{PucPaoGio:J22,FavMatPuc:22} can be obtained by
\begin{equation}
\mathbf{w}_\mathrm{T} = \sqrt{\rho_\mathrm{p}}\,\mathbf{w}_\mathrm{T,s} + \sqrt{1-\rho_\mathrm{p}}\,\mathbf{w}_\mathrm{T,c}
\label{BF vector}
\end{equation}
where $\rho_\mathrm{p} \in [0,1]$ represents the fraction of power reserved for sensing in \ac{JSC} slots, thus determining the fraction of power dedicated to communication, i.e., $1-\rho_\mathrm{p}$, while $\mathbf{w}_\mathrm{T,s}$ and $\mathbf{w}_\mathrm{T,c}$ are respectively the sensing and communication beamforming vectors. 
By considering a beam steering approach and performing a normalization with respect to the \ac{EIRP} $P_\mathrm{T}G_\mathrm{T}^\mathrm{a}$, the latter are  
\begin{equation*}
\label{eq:tx_bf_vector}
\mathbf{w}_\mathrm{T,s} = \frac{\sqrt{P_\mathrm{T} G_\mathrm{T}^\mathrm{a}}}{N_\mathrm{T}}\mathbf{a}_\mathrm{T}^{*}(\theta_\mathrm{T,s}), \,\,\,
\mathbf{w}_\mathrm{T,c} = \frac{\sqrt{P_\mathrm{T} G_\mathrm{T}^\mathrm{a}}}{N_\mathrm{T}}\mathbf{a}_\mathrm{T}^{*}(\theta_\mathrm{T,c})
\end{equation*}
where $P_\mathrm{T}$ is the transmit power, $G_\mathrm{T}^\mathrm{a}$ is the transmit array gain along the beam steering direction, and $\mathbf{a}_\mathrm{T}(\theta_\mathrm{T,s}) \in \mathbb{C}^{N_\mathrm{T} \times 1}$ and $\mathbf{a}_\mathrm{T}(\theta_\mathrm{T,c}) \in \mathbb{C}^{N_\mathrm{T} \times 1}$ are the steering vectors for sensing and communication directions, $\theta_\mathrm{T,s}$ and $\theta_\mathrm{T,c}$, respectively \cite{PucPaoGio:J22}.

Fig.~\ref{fig:block} provides a graphical representation of the resource allocation for communication and \ac{JSC} sensing and an example of the evolution of these resources over time. It is noteworthy that within the time-frequency slots designated for \ac{JSC}, two scenarios can unfold: a) the slot is solely dedicated to sensing ($\rho_\mathrm{p}=1$); b) the slot accommodates one concurrent \ac{UE} (i.e., one downlink communication beam), sharing resources with the sensing task according to \eqref{BF vector}.

At the sensing receiver, the symbols after \ac{OFDM} demodulation are represented by
%
\begin{equation}
\mathbf{y}[k,m] = \mathbf{H}[k,m] \mathbf{x}[k,m] + \mathbf{n}[k,m] 
    \label{eq:y_tilde}
\end{equation}
where $\mathbf{H}[k,m] \in \mathbb{C}^{N_\mathrm{R} \times N_\mathrm{T}}$ is the channel matrix for the $m$th symbol and the $k$th subcarrier, and $\mathbf{n}[k,m] \sim \mathcal{CN}(\mathbf{0},\sigma_\mathrm{N}^2 \mathbf{I}_{N_\mathrm{R}})$ is the noise vector at the receiving antennas.
In the considered sensing direction, denoted as $\theta_\mathrm{R,s}=\theta_\mathrm{T,s}$, spatial combining is performed using the receiving beamforming vector $\mathbf{w}_\mathrm{R} = \mathbf{a}_\mathrm{R}^*(\theta_\mathrm{R,s})$. This process results in the formation of a grid of received symbols denoted as $\mathbf{Y} \in \mathbb{C}^{K_\mathrm{s} \times M_\mathrm{s}}$, where each element $y_{k,m}$ is obtained by taking the inner product between the receiving beamforming vector $\mathbf{w}_\mathrm{R}$ and the vector of the symbols received at each antenna $\mathbf{y}[k,m]$, i.e., $y_{k,m} = \mathbf{w}_\mathrm{R}^\mathsf{T} \mathbf{y}[k,m]$.
The collected symbols are subsequently used to generate range-angle maps as illustrated in Section~\ref{sec:rangeangle}. 

The spatial steering vectors $\mathbf{a}_\mathrm{T}(\theta)$ and $\mathbf{a}_\mathrm{R}(\theta)$ for a \ac{ULA} with half-wavelength interelement spacing at a given \ac{AoA}/\ac{AoD} $\theta$ is given by \cite[Chapter~9]{richards},\cite[Chapter~5]{asplund2020advanced}
\begin{equation*}
\mathbf{a}_\mathrm{T,R}(\theta)=\left [1, e^{\imath \pi \sin(\theta)}, \dots, e^{\imath \pi (N_\mathrm{a}-1) \sin(\theta)} \right]^\mathsf{T} 
\label{eq:steering2}
\end{equation*}
where $N_\mathrm{a}$ is the number of array antenna elements.

%

\subsection{Multipath Channel Model}\label{sec:channelMulti}
Let us begin by considering an anisotropic point-like scatterer labeled as $l$. This scatterer can either be a point-like target or a reflection point of a more complex object, referred to as an extended target. The channel model experienced during sensing is illustrated in Fig.~\ref{fig:multipath}. The signal emitted by the $j$th antenna propagates through the forward channel, from the \ac{BS} to the scatterer, and is then reflected by the scatterer, propagating back to the $i$th antenna at the \ac{BS} via the backscatter channel.

In the presence of the sole direct path in both forward and backscatter channels, the linear time-variant \ac{MIMO} channel\footnote{The time-variant nature accounts for the Doppler effect due to target motion.} is represented in the frequency domain by the channel matrix $\mathbf{H}[k,m]$ in \eqref{eq:y_tilde}, which takes the form \cite{Fur:J21a,Fur:J21b,PucPaoGio:J22}
\begin{equation*}\label{eq:channel-mat-directonly}
    \mathbf{H}_l[k,m] =  \nu_l(\psi_l)\alpha_l e^{\imath \phi_l}e^{\imath 2\pi m T_\mathrm{s} f_{\mathrm{D},l}}e^{-\imath 2\pi k \Delta f \tau_l} \mathbf{a}_\mathrm{R}(\theta_l)\mathbf{a}^\mathsf{T}_\mathrm{T}(\theta_l)
\end{equation*}
where $\tau_l$, $f_{\mathrm{D},l}$, and $\theta_l$ 
are the round-trip delay, Doppler shift, and \ac{AoA}/\ac{AoD} of the $i$th (direct) Tx-scatterer-Rx path, respectively, while $\alpha_l\in \mathbb{R}$ is the gain which accounts for the path-loss (described by the radar equation as this is the sensing channel) and the \ac{RCS} of the  scatterer, and $\phi_l$ is a random phase.
The function $\nu_l(\psi_l)$ capturing the scatterer anisotropy, is in the following dubbed visibility function, which accounts for angular-dependent reflection properties, where $\psi_l$ is the \ac{AoA} at the target side for the \ac{BS}-scatterer link. Details will be provided in Section~\ref{Sec:target}. 
In the presence of multipath propagation represented by $L_\mathrm{p}$ paths between the \ac{BS} and the scatterer, it can be shown that, due to linearity, the channel matrix in \eqref{eq:y_tilde} becomes\footnote{An interesting interpretation provided in \cite{alhassoun2019theoretical, blunt2011performance} is that the overall \ac{CIR} can be represented as the convolution between two \acp{CIR}: one from the so-called forward channel and the other from the backscatter channel. These \acp{CIR} are scaled by a factor that accounts for the path-loss and \ac{RCS} of the target.} 
\begin{align}\label{eq:channel-mat-OneReflectinitial}
    \mathbf{H}_l[k,m] & = \sum_{p = 0}^{L_\mathrm{p}-1} \sum_{q = 0}^{L_\mathrm{p}-1} \nu_l(\psi_l^p)\nu_l(\psi_l^q)\xi^p_l \xi^q_l e^{\imath (\phi^p_l+\phi^q_l)}\\
    & \times e^{\imath 2\pi m T_\mathrm{s} (f^p_{\mathrm{D},l}+f^q_{\mathrm{D},l})} e^{-\imath 2\pi k \Delta f (\tau^p_l+\tau^q_l)} \mathbf{a}_\mathrm{R}(\theta^p_l)\mathbf{a}^\mathsf{T}_\mathrm{T}(\theta^q_l)\nonumber
\end{align}
where $\xi^r_l$, $\phi^r_l$, $\tau^r_l$, $f^r_{\mathrm{D},l}$, and $\theta^r_l$ 
are the amplitude, phase, delay, Doppler shift, and \ac{AoA}/\ac{AoD} of the $r$th path (at the \ac{BS}), respectively, while $\psi_l^r$ are the angles at the scatterer side. The channel matrix \eqref{eq:channel-mat-OneReflectinitial} can be conveniently reformulated by separating the \ac{BS}-scatterer-\ac{BS} path (the one with $p=q=0$), with  $(\xi_l^0)^2=\alpha_l$, from the diffuse 
component resulting into

\begin{align}\label{eq:channel-mat-OneReflect}
    \mathbf{H}_l&[k,m] = \underbrace{\nu_l(\psi_l)\alpha_l \, e^{\imath\phi_l}e^{\imath 2\pi m T_\mathrm{s} f_{\mathrm{D},l}}e^{-\imath 2\pi k \Delta f \tau_l} \mathbf{a}_\mathrm{R}(\theta_l)\mathbf{a}^\mathsf{T}_\mathrm{T}(\theta_l)}_{\textrm{direct path}}\nonumber\\
    & \quad + 
    \sum_{p = 1}^{L_\mathrm{p}-1} \sum_{q = 1}^{L_\mathrm{p}-1} \nu_l(\psi_l^p)\nu_l(\psi_l^q) \xi^p_l \xi^q_l 
    e^{\imath (\phi^p_l+\phi^q_l)}\\
    & \quad \underbrace{\times e^{\imath 2\pi m T_\mathrm{s} (f^p_{\mathrm{D},l}+f^q_{\mathrm{D},l})} e^{-\imath 2\pi k \Delta f (\tau^p_l+\tau^q_l)} \mathbf{a}_\mathrm{R}(\theta^p_l)\mathbf{a}^\mathsf{T}_\mathrm{T}(\theta^q_l)}_{\textrm{ diffuse component}}.\nonumber
\end{align}

Note that the diffuse component in \eqref{eq:channel-mat-OneReflect} vanishes in the absence of multipath. Typically, the diffuse part remains unknown to the sensing receiver; thus, it can act as a disturbance, causing target smearing in the range-angle maps, as it will be shown in Section~\ref{sec:results}. Due to multiple paths, even a single scatterer is sensed as a superposition of various components with distinct delays, Doppler shifts, and \acp{AoA}, spreading the target across all three domains.

In the presence of $L_\mathrm{s}$ point-like targets or a single extended target modeled as a collection of $L_\mathrm{s}$ scatterers, the channel matrix becomes
\begin{align}\label{eq:channel-mat-LsReflect}
    \mathbf{H}[k,m] = \sum_{l=1}^{L_\mathrm{s}}\mathbf{H}_l[k,m] 
\end{align}
which contains $L_\mathrm{s}$ direct paths plus $L_\mathrm{s}(L_\mathrm{p}^2-1)$ diffuse paths.

%

\subsection{Target Model}\label{Sec:target}
In this work, we consider a scenario featuring cars represented by the extended target model depicted in Fig.~\ref{fig:multipath} and pedestrians as point-like targets.
The channel matrix outlined in \eqref{eq:channel-mat-OneReflect} and \eqref{eq:channel-mat-LsReflect} accounts for the reflections caused by each scatterer, resulting in the generation of one or more backscattered signals. 
For each scatterer, the gain $\alpha_l$ includes the attenuation along the BS-scatterer-BS path, which can be related to the radar equation, and results in \cite{richards}
\begin{equation*}
\alpha_l = \sqrt{\frac{c^2 \sigma_l}{(4\pi)^3 f_\mathrm{c}^2 d_l^4}}
\label{eq:amplitude}
\end{equation*}
where $d_l$ is the distance between the $l$th reflection point and the considered \ac{BS}, and $\sigma_l$ is its \ac{RCS}, while $f_\mathrm{c}$ and $c$ represent the carrier frequency and the speed of light, respectively. The scatterers adhere to the Swerling~I model, i.e., with \ac{p.d.f.} given by \cite{Sko:B08}
\begin{equation}
    f(\sigma_l) = \frac{1}{\bar{\sigma}_l}\exp \left(-\frac{\sigma_l}{\bar{\sigma}_l}\right) \qquad \sigma_l \geq 0
    \label{SwerI}
\end{equation}
which is constant during the collection of a block of $M_\mathrm{s}$ symbols, and independent from block to block, with $\bar{\sigma}_l=\mathbb{E}\{\sigma_l\}$.

The pedestrian is modeled as a point-like target with $\bar{\sigma}_l = 0\,\text{dBms}$. The car is an extended target consisting of $L_\mathrm{s}=12$ reflection points, including $4$ reflections attributed to the front, back, and sides, each with a large \ac{RCS} ($\bar{\sigma}_l = 20\,\text{dBms}$). Additionally, there are $4$ reflections associated with the wheelhouses ($\bar{\sigma}_l = 0\,\text{dBms}$) and $4$ reflections originating from the corners ($\bar{\sigma}_l = 5\,\text{dBms}$) \cite{buhren:06Automotive, BuhrenYang:06}. Each scatterer is anisotropic, and its angle-dependent reflectivity is modeled through the visibility function $\nu_l(\psi_l)$, which characterizes the angular range where the scatterer exhibits an appreciable \ac{RCS}; in contrast, the \ac{RCS} is negligible outside that angular range.
Recalling that $\psi_l$ is the \ac{AoA} at the target side for the \ac{BS}-scatter link, for the car, $\nu_l(\psi_l)=1$ for $\psi_l\in [\Psi_{\mathrm{min},l},\Psi_{\mathrm{max},l}]$ and $0$ elsewhere, with $\Psi_{\mathrm{min},l}$ and $\Psi_{\mathrm{max},l}$ defined for the specific scatterer type (e.g., corner, wheelhouse, etc.) \cite{buhren:06Automotive, BuhrenYang:06}. The visibility functions adopted in this work are shown in Fig.~\ref{fig:multipath}.

Note that the number of back-scattered signals varies over time because the number of reflections generated by the extended target depends on its position and orientation for the considered \ac{BS}, according to the visibility function. Each scatterer, when visible, is associated with the \ac{RCS} model in \eqref{SwerI}. Channels experienced by different scatterers are considered statistically independent; such an assumption is plausible because the scattering points of the extended target are spaced apart by several wavelengths at mmWave. 

%
%

\subsection{Range-Angle Map at the Base Station}\label{sec:rangeangle}
To detect objects within the environment, each \ac{BS} uses a combination of two beams specified by $\mathbf{w}_\mathrm{T}$ in \eqref{BF vector}. The communication beam is aimed at a particular \ac{UE}, while the sensing beam is periodically steered within the range $[-\Theta,\Theta]$ with a fixed angular increment $\Delta \Theta$. A group of $M_\mathrm{s}$ \ac{OFDM} symbols is transmitted to produce a range-angle map for every sensing direction. The duration of a full scan, referred to as $T_\mathrm{scan}$, is determined by the number of sensing orientations chosen and the duration of each \ac{OFDM} symbol, $T_\mathrm{s}$. After collecting all symbols in a given scan direction in the matrix $\mathbf{Y}$, the first stage is to perform reciprocal filtering, which consists of an element-wise division between the received and the transmitted grids, i.e., $\mathbf{Y}$ and $\mathbf{X}$, to remove the dependence on the transmitted symbols and obtain a new matrix $\mathbf{G}$ whose elements are \cite{Braun,RodBlaColLom:J23,PucPaoGio:J22}
\begin{equation*}
    g_{k,m}=\frac{y_{k,m}}{x_{k,m}}=\mathbf{w}_\mathrm{R}^\mathsf{T}\mathbf{H}[k,m]\mathbf{w}_\mathrm{T}+\frac{\mathbf{w}_\mathrm{R}^\mathsf{T}\mathbf{n}[k,m]}{x_{k,m}}.
\end{equation*}

Then, a double-periodogram is computed on the rows and columns of $\mathbf{G}$ to obtain a range-Doppler map, following the approach outlined in \cite{Braun,FullDuplex,PucPaoGio:J22}. Given the relatively narrow beamwidth, we assume that only one scatterer will likely be present in each sensing direction. Therefore, the range-angle map is calculated by selecting the column of the periodogram with the maximum value, thereby associating it uniquely with the corresponding sensing direction.\footnote{This is a low-complexity solution that can be readily replaced with alternative approaches such as 2D-MUSIC \cite{HenManArnBri:C22}.}

\subsection{Network Capacity and Communication Overhead}
The dual functional capability of the \ac{BS} is ensured by considering frequency-division outlined in \eqref{eq:rho_f}, time-division introduced in \eqref{eq:rho_t}, and power-division reported in \eqref{BF vector}. 

The assignment of resources to sensing inevitably influences communication performance. To assess this impact, we consider the aggregate capacity, defined as the downlink sum rate of each \ac{BS}, calculated as\footnote{To streamline the scenario we consider that each \ac{BS} can use all the available resources.}
\begin{align}\label{eq:Shannon}
C_\mathrm{DL}\left(\rho_{\mathrm{f}},\rho_{\mathrm{t}} ,\rho_{\mathrm{p}} \right) &= \rho_\mathrm{t} \, \Delta f \sum_{k=1}^{\lfloor\rho_{\mathrm{f}}K_0\rceil} C_\mathrm{JSC}\left( (1-\rho_\mathrm{p})\mathrm{SNR}^{(k)}_\mathrm{c}\right)\nonumber\\
   & +\rho_\mathrm{t} \, \Delta f \sum_{k=\lfloor\rho_{\mathrm{f}}K_0+1\rceil}^{K_0} \log_2(1+\mathrm{SNR}^{(k)}_\mathrm{c})\nonumber\\
   & +(1-\rho_\mathrm{t})\Delta f \sum_{k=1}^{K_0} \log_2(1+\mathrm{SNR}^{(k)}_\mathrm{c})
\end{align}
where $\mathrm{SNR}^{(k)}_\mathrm{c}$ is the communication \ac{SNR} experienced by the user at subcarrier $k$ and $C_\mathrm{JSC}\left(\gamma\right)$ is the capacity of the \ac{JSC} slots. Indeed, while slots allocated for communication may use any constellation and the sum rate for them is calculated via the usual Shannon capacity, the \ac{JSC} slots utilize a fixed constellation (in this case, \ac{QPSK}) as it ensures high sensing performance because of its constant modulus \cite{XioLiu:J23}.\footnote{For \ac{QPSK} signaling, a tight approximation of the capacity can be found in \cite{ChoSheHon:19}:
\begin{equation*}
    C_\mathrm{JSC}(\gamma)=2(1-a_1e^{-b_1\gamma}-a_2e^{-b_2\gamma})
\end{equation*}
where $\gamma$ is the \ac{SNR}, $a_1 = 0.143281$, $a_2 = 0.856719$, $b_1 = 1.557531$, and $b_2 = 0.57239$.
} 


Sensing not only consumes resources at the radio segment, but cooperation among the \ac{BS}, as illustrated in the next section, requires exchanging information via the backhaul contributing to network overhead. The overhead generated by the range-angle maps transmitted to the \ac{FC} can be quantified as the average bit rate (averaged over the cooperating \acp{BS})
\begin{equation}\label{eq:overhead}
    R_\mathrm{b} =  \frac{ N_\mathrm{b} \rho_\mathrm{t} }{T_{\mathrm{meas}}} \frac{1}{N_\mathrm{s}}\sum_{s=1}^{N_\mathrm{s}} N_\mathrm{p}^{(s)}
\end{equation}
where $N_\mathrm{s}$ is the number of \acp{BS}, $N_\mathrm{p}^{(s)}$ represents the number of map points shared by the $s$th \ac{BS} and $N_\mathrm{b}$ is the number of bits used to encode each map point.


\section{Multi-Sensor Data Fusion and Tracking}\label{sec:tracking}

\subsection{Maps Fusion from Multiple Cooperating \acp{BS}}
\begin{figure*}[t]
    \centering
    \includegraphics[width=2\columnwidth]{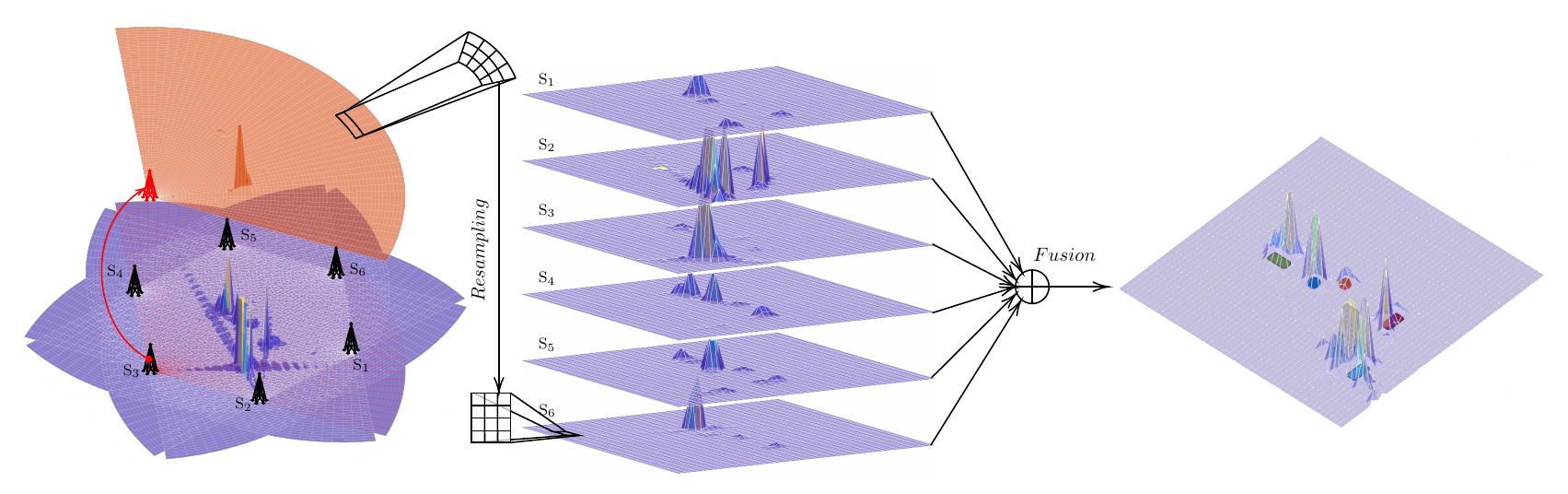}
    \caption{Soft map fusion strategy. The maps shown are obtained from the simulations reported in the numerical results considering the vehicle scenario described in Section~\ref{sec:results}.}
    \label{fig:maps_fusion}
\end{figure*}

The strategy implemented to process the range-angle maps obtained by the network of \ac{JSC} \acp{BS} is depicted in Fig.~\ref{fig:block_diagram}, and the map fusion sketched in Fig.~\ref{fig:maps_fusion}. Each \ac{BS} performs a uniform resampling procedure with grid resolution $\Delta_\mathrm{x}$ and $\Delta_\mathrm{y}$ to ensure consistent map fusion. Then, an excision filter with threshold $\gamma_\mathrm{s}$ is used to sift relevant map points (those related to regions where the score, or map intensity, is high) and reduce the network overhead due to map exchange with the \ac{FC}. Hence, only selected pixels are shared with the \ac{FC}. The effect of map excision and the related sensing/overhead trade-off will be addressed in Section~\ref{sec:results}.

At the \ac{FC}, the resampled and filtered range-angle maps $\mathbf{L}_{s,t}$ are then combined using element-wise summation to obtain the soft map $\mathbf{L}_{t} = \frac{1}{N_\mathrm{s}}\sum_{s=1}^{N_\mathrm{s}}\mathbf{L}_{s,t}$, where $t$ is the time index.

\subsection{Clustering}\label{sec:clustering}
\begin{figure}[t]
    \centering
    \includegraphics[width=\columnwidth]{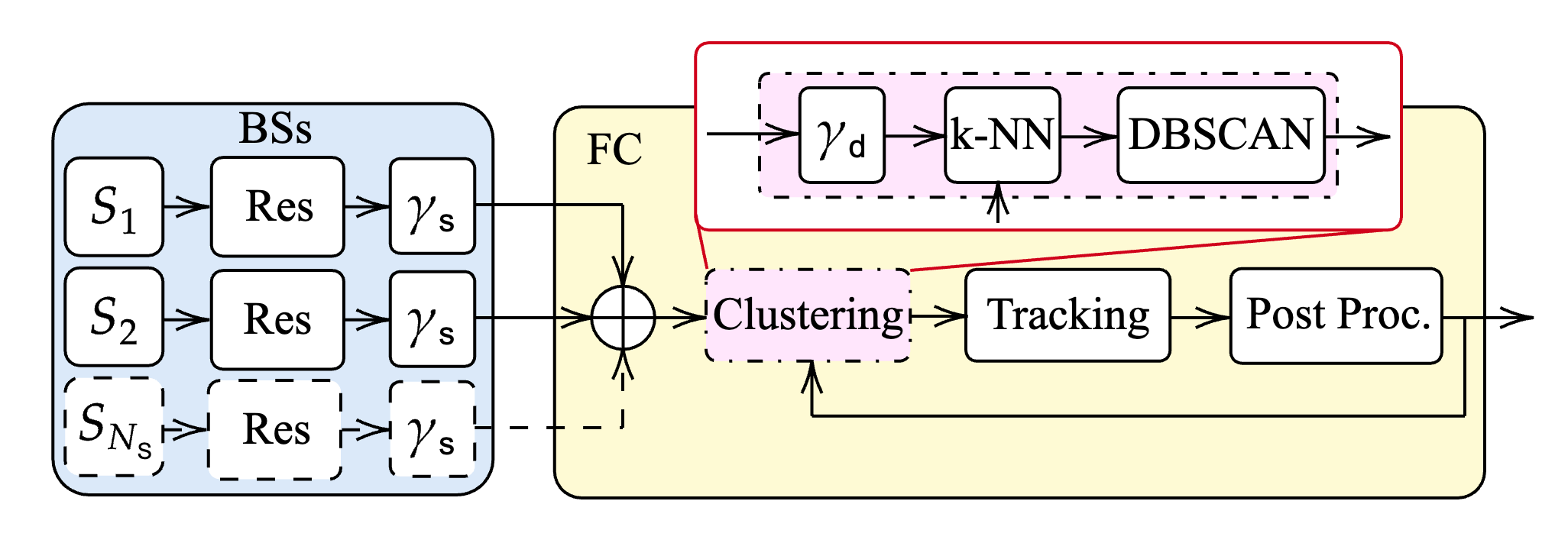}
    \caption{Block diagram of the processing chain. \acp{BS} and \ac{FC} are depicted in blue and yellow, respectively.}
    \label{fig:block_diagram}
    \label{fig:cluster_diagram}
\end{figure}

The three-step clustering procedure used for performing detections from the soft maps and handling extended objects is illustrated in Fig.~\ref{fig:cluster_diagram}. In the first step, an excision filter is applied to the matrix $\mathbf{L}_{t}$ with a threshold $\gamma_\mathrm{d}$ to retain map points with higher scores.
In the second step, a \ac{k-NN} algorithm with $k=1$ and a gate parameter $\xi_\mathrm{NN}$ is used to cluster points that are likely associated with previously detected targets (as discussed in \cite{WatBorKat:16}).
The third step involves clustering of the remaining points, which are map points exceeding $\gamma_\mathrm{d}$ and lying outside the gate $\xi_\mathrm{NN}$ using the \ac{DBSCAN} algorithm. This algorithm specifies a maximum distance $\xi_\mathrm{d}$ for points to be considered part of the same cluster, as well as a minimum number of points, denoted as $N_\mathrm{d}$, required to form a cluster (as described in \cite{EstKriXia:96}). It is important to note that \ac{DBSCAN} is utilized in this context to cluster residual points that can represent potential new spawning targets.
Lastly, each cluster centroid is stored in the matrix $\mathbf{Z}_{t}$, representing the target detections extracted from the soft maps.

\subsection{Multi-target Tracking}
In this section, we introduce the algorithms used for multi-target tracking, namely the \ac{PHD} filter and the \ac{MBM} filter.
The following state vector represents each target state
\begin{equation*}
    \mathbf{s}_{t,n} = \big(s_{t,n,\mathrm{x}},  s_{t,n,\mathrm{y}},  s_{t,n,v_\mathrm{x}},  s_{t,n,v_\mathrm{y}}\big)^\mathsf{T}
\end{equation*}
where $n$ is the target index. The first two elements of this vector correspond to the target position coordinates, while the last two represent the target velocity components. In this work, we update the target position components using information extracted from the map and calculate the target velocity by considering both the previous target position (at time $t-1$) and the current position.

\subsubsection{Probability Hypothesis Density Filter}
The \ac{PHD} filter is a widely known algorithm for tracking \cite{Mah:03}. A typical implementation approximates the target intensity function as a \ac{GM} as \cite{VoMa:06} 
\begin{equation}
    D_{t-1|t-1}(\mathbf{x}) =\!\!\!\! \sum_{h=1}^{\mathcal{H}_{t-1|t-1}} \!\!\! w_{t-1|t-1}^{(h)} \mathcal{N}_\mathbf{x} (\boldsymbol{\mu}_{t-1|t-1}^{(h)},\mathbf{P}_{t-1|t-1}^{(h)})
    \label{eq:PHD}
\end{equation}
where $\mathbf{x}$ is a \ac{RFS}, $\mathcal{H}_{t-1|t-1}$ represents the number of Gaussian components in the prior intensity function, while $w_{t-1|t-1}^{(h)}$, $\boldsymbol{\mu}_{t-1|t-1}^{(h)}$, and $\mathbf{P}_{t-1|t-1}^{(h)}$, represent importance, mean and covariance of the $h$th component.
The intensity function can be interpreted as a non-normalized \ac{p.d.f.} whose integral returns the estimated number of targets in the scenario.

Linear Kalman prediction is used to infer the intensity function, i.e., $D_{t|t-1}(\mathbf{x})$, which can still be written in the form of equation \eqref{eq:PHD}, with parameters 
\begin{align}\label{eq:kalman}
    w_{t|t-1}^{(h)} &= P_\mathrm{s}  w_{t-1|t-1}^{(h)} \nonumber\\
    \boldsymbol{\mu}_{t|t-1}^{(h)} &= \mathbf{F} \boldsymbol{\mu}_{t-1|t-1}^{(h)} \nonumber\\
    \mathbf{P}_{t|t-1}^{(h)} &= \mathbf{F} \mathbf{P}_{t-1|t-1}^{(h)} \mathbf{F}^\mathsf{T} + \mathbf{Q}
\end{align}
where $P_\mathrm{s}$ stands for the probability of survival of the target, $\mathbf{F}$ represents the transition matrix that models the target behavior, and $\mathbf{Q}$ is the covariance matrix that represents the motion uncertainty.

To build the predicted intensity function, the $\mathcal{B}$ birth components must be added, representing the possibility of new targets spawning in the surveillance area. The total number of components after prediction is then $\mathcal{H}_{t|t-1} = \mathcal{H}_{t-1|t-1} + \mathcal{B}$.

The predicted components are then updated with the measurements $\mathbf{Z}_t$ extracted from the map $\mathbf{L}_t$, as described in Section~\ref{sec:clustering}. Firstly, the Kalman filter parameters are derived as
\begin{align}\label{eq:update1} 
        \hat{\mathbf{z}}_{t|t-1}^{(h)} &= \mathbf{H} \boldsymbol{\mu}_{t|t-1}^{(h)} \nonumber\\
        \mathbf{S}_{t,m}^{(h)} &= \mathbf{R}_{t,m} + \mathbf{H} \mathbf{P}^{(h)}_{t|t-1} \mathbf{H}^\mathsf{T} \nonumber\\
        \mathbf{K}_{t,m}^{(h)} &= \mathbf{P}^{(h)}_{t|t-1} \mathbf{H}^\mathsf{T} (\mathbf{S}_{t,m}^{(h)})^{-1}    
\end{align}
where $\mathbf{H}$ represents the measurement matrix, $\hat{\mathbf{z}}_{t|t-1}^{(h)}$ is the $h$th predicted measurement, $\mathbf{S}_{t,m}^{(h)}$ is the innovation covariance, $\mathbf{R}_{t,m}$ represents the $m$th measurement covariance, and $\mathbf{K}_{t,m}^{(h)}$ is the Kalman gain. Then, each component in the predicted intensity function is updated with each measurement $\mathbf{z}_{t,m}$ in the measurements matrix $\mathbf{Z}_t$, by
\begin{align*}
    \widetilde{w}_{t|t}^{(j)} &= P_\mathrm{d}  w_{t|t-1}^{(h)} \mathcal{N}_{\mathbf{z}_{t,m}}(\hat{\mathbf{z}}_{t|t-1}^{(h)},\mathbf{S}_{t,m}^{(h)})\\
    \boldsymbol{\mu}_{t|t}^{(j)} &= \boldsymbol{\mu}_{t|t-1}^{(h)} +  \mathbf{K}_{t,m}^{(h)} (\mathbf{z}_{t,m} - \hat{\mathbf{z}}^{(h)}_{t|t-1})\\
    \mathbf{P}^{(j)}_{t|t} &= (\mathbf{I} - \mathbf{K}_{t,m}^{(h)} \mathbf{H}) \mathbf{P}^{(h)}_{t|t-1}
\end{align*}
where $P_\mathrm{d}$ represents the probability of detection, and $\mathcal{N}_{\mathbf{z}_{t,m}}(\hat{\mathbf{z}}_{t|t-1}^{(h)},\mathbf{S}_{t,m}^{(h)})$ stands for the $m$th measurement likelihood with respect to the $h$th component.
Weights are then normalized as
\begin{equation*}
    w_{t|t}^{(j)} = \frac{\widetilde{w}_{t|t}^{(j)}}{\lambda_c + \sum_{k=1}^{\mathcal{H}_{t|t-1}} \widetilde{w}_{t|t}^{(k)}}
\end{equation*}
where $\lambda_c$ represents the clutter intensity, which is modelled as a \ac{PPP} \cite{GarXiaGra:19}. To account for missed detections, an additional set of components is added to the posterior, i.e.,
\begin{align*}
    w_{t|t}^{(h)} &= (1-P_\mathrm{d})  w_{t|t-1}^{(h)}, \,\,
    \boldsymbol{\mu}_{t|t}^{(h)} &= \boldsymbol{\mu}_{t|t-1}^{(h)}, \,\,
    \mathbf{P}^{(h)}_{t|t} &= \mathbf{P}^{(h)}_{t|t-1}.
\end{align*}

The number of components in the posterior can be written as $\mathcal{H}_{t|t} = \mathcal{H}_{t|t-1}(M_t + 1)$, where $M_t$ stands for the number of measurements at time instant $t$. Hence, the number of targets is estimated via
\begin{equation*}
    \widehat{N}_{\mathrm{obj}} = \nint*{\sum_{h=1}^{\mathcal{H}_{t|t}} w_{t|t}^{(h)}}
\end{equation*}
while the $n$th target state estimation is obtained by
\begin{equation*}
    \hat{\mathbf{s}}_{t,n} = \argmax_{ w_{t|t}^{(h)}} \boldsymbol{\mu}_{t|t}^{(h)}.
\end{equation*}

\subsubsection{Multi-Bernoulli Mixture Filter}
The \ac{MBM} filter is an alternative to the \ac{PHD} filter in multiple target tracking problems that exploits the association probability between measurements and targets \cite{GarWilGra:18}. Let us consider the following \ac{MBM} distribution to represent the prior multiobject \ac{p.d.f.}
\begin{equation}
    \mathrm{MBM}_{t-1|t-1}(\mathbf{x}) = \sum_{g = 1}^{\mathcal{G}_{t-1|t-1}} w^{(g)}_{t-1|t-1} \mathrm{MB}^{(g)}_{t-1|t-1}(\mathbf{x})
    \label{eq:MBM}
\end{equation}
where $\mathcal{G}_{t-1|t-1}$ represents the number of \ac{MB} components or global hypotheses and $w^{(g)}_{t-1|t-1}$ stands for the $g$th component importance. 
The \ac{MB} distribution \eqref{eq:MBM} can be written as 
\begin{equation}
   \mathrm{MB}^{(g)}_{t-1|t-1}(\mathbf{x}) = \sum_{\biguplus \mathbf{x}_l = \mathbf{x}} \prod_{l = 1}^{\mathcal{L}^{(g)}_{t-1|t-1}}  \mathrm{B}^{(g,l)}_{t-1|t-1}(\mathbf{x}_l)
   \label{eq:MB}
\end{equation}
where $\mathcal{L}^{(g)}_{t-1|t-1}$ is the number of Bernoulli components or local hypotheses, and the summation is performed over all possible unions of mutually disjoint \acp{RFS} that generate $\mathbf{x}$, i.e., all possible data associations between measurements and targets \cite{GarWilGra:18}. 
Each component in \eqref{eq:MB} can be written as
\begin{equation*}
    \mathrm{B}^{(g,l)}_{t-1|t-1}(\mathbf{x}_l) = r_{t-1|t-1}^{(g,l)} \mathcal{N}_{\mathbf{x}_l}(\boldsymbol{\mu}_{t-1|t-1}^{(g,l)},\mathbf{P}_{t-1|t-1}^{(g,l)})
\end{equation*}
where $r_{t-1|t-1}^{(g,l)}$ is the existence probability of the $l$th local hypothesis in the $g$th global hypothesis, while $\boldsymbol{\mu}_{t-1|t-1}^{(h)}$ and $\mathbf{P}_{t-1|t-1}^{(h)}$ represent the mean and the covariance of the considered component, respectively.
Overall, the \ac{MBM} distribution can be represented with the following set of parameters:
\begin{equation*}
    \bigl\{w_{t-1|t-1}^{(g)}, \bigl\{r_{t-1|t-1}^{(g,l)}, \boldsymbol{\mu}_{t-1|t-1}^{(g,l)}, \mathbf{P}_{t-1|t-1}^{(g,l)}\bigr\}_{l=1}^{\mathcal{L}^{(g)}_{t-1|t-1}} \bigr\}_{g=1}^{\mathcal{G}_{t-1|t-1}}.
\end{equation*}
Similarly to \eqref{eq:kalman} a Kalman prediction is performed to infer the parameters at time $t$ as follows
\begin{align}\label{eq:kalman2}
    w_{t|t-1}^{(g)} &= w_{t-1|t-1}^{(g)},\,\,\,
    r_{t|t-1}^{(g,l)} = P_\mathrm{s}  r_{t-1|t-1}^{(g,l)} \nonumber\\
    \boldsymbol{\mu}_{t|t-1}^{(g,l)} &= \mathbf{F} \boldsymbol{\mu}_{t-1|t-1}^{(g,l)},\,\,\,
    \mathbf{P}_{t|t-1}^{(g,l)} = \mathbf{F} \mathbf{P}_{t-1|t-1}^{(g,l)} \mathbf{F}^\mathsf{T} + \mathbf{Q}.
\end{align}
To account for new spawning targets, a set of $\mathcal{B}$ Bernoulli components is added to each global hypothesis. The overall number of components after the prediction step is $(\mathcal{L}_{t-1|t-1}+\mathcal{B})\mathcal{G}_{t-1|t-1}$. In the update phase, firstly, Kalman filter parameters are derived, i.e.,
\begin{align}\label{eq:update2}
        \hat{\mathbf{z}}_{t|t-1}^{(g,l)} &= \mathbf{H} \boldsymbol{\mu}_{t|t-1}^{(g,l)} \nonumber\\
        \mathbf{S}_{t,m}^{(g,l)} &= \mathbf{R}_{t,m} + \mathbf{H} \mathbf{P}^{(g,l)}_{t|t-1} \mathbf{H}^\mathsf{T} \nonumber\\
        \mathbf{K}_{t,m}^{(g,l)} &= \mathbf{P}^{(g,l)}_{t|t-1} \mathbf{H}^\mathsf{T} (\mathbf{S}_{t,m}^{(g,l)})^{-1}. 
\end{align}
Then, each component in the predicted \ac{MBM} distribution is updated with measurements. 
Let us define the association vector $\boldsymbol{\theta}_i$ for the $g$th global hypothesis, whose elements $\theta_{i,j} = m$ represent the association between the $l$th measurement and the $m$th Bernoulli component. 
If a hypothesis is associated with a missed detection, its association vector elements $\theta_{i,j}$ is set to $0$.
Each association hypothesis represents a new component in the \ac{MBM} posterior.
During update, if a component is associated with a measurement, i.e., $\theta_{i,j} \neq 0$, that component parameters are updated according to
\begin{align*}
    r_{t|t}^{(i,j)} &= 1 \\
    \boldsymbol{\mu}_{t|t}^{(i,j)} &= \boldsymbol{\mu}_{t|t-1}^{(h)} +  \mathbf{K}_{t,m}^{(g,l)} (\mathbf{z}_{t,m} - \hat{\mathbf{z}}^{(g,l)}_{t|t-1})\\
    \mathbf{P}^{(i,j)}_{t|t} &= (\mathbf{I} - \mathbf{K}_{t,m}^{(g,l)} \mathbf{H}) \mathbf{P}^{(g,l)}_{t|t-1}.
\end{align*}
Conversely, if a component is associated to a missed detection, i.e., $\theta_{i,j} = 0$, then the parameters of this component are updated as follows
\begin{align*}
    r_{t|t}^{(i,j)} &= \frac{r_{t|t-1}^{(g,l)}(1-P_\mathrm{d})}{1-r_{t|t-1}^{(g,l)}+r_{t|t-1}^{(g,l)}(1-P_\mathrm{d})} \\
    \boldsymbol{\mu}_{t|t}^{(i,j)} &= \boldsymbol{\mu}_{t|t-1}^{(g,l)} \\
    \mathbf{P}_{t|t}^{(i,j)} &= \mathbf{P}_{t|t-1}^{(g,l)}.
\end{align*}
It is a common practice to define log-weights $l_{t|t-1}^{(g)} = \log(w_{t|t-1}^{(g)})$ to avoid numerical problems in the posterior weight computation \cite{ThaAleRub:21}.
For a given measurement association $\theta_{i,j} = m$ in the association vector, we can compute its posterior log-weight as
\begin{equation*}
   l_{t|t-1}^{(i,j)} = \log(r_{t|t-1}^{(g,l)} P_\mathrm{d} \lambda_c^{-1} \mathcal{N}_{\mathbf{z}_{t,m}}(\mathbf{H} \boldsymbol{\mu}_{t|t-1}^{(g,l)}, \mathbf{S}_{t,m}^{(g,l)}))
\end{equation*}
and in case of missed detection, $\theta_{i,j} = 0$, as
\begin{equation*}
   l_{t|t-1}^{(i,j)} = \log(1-r_{t|t-1}^{(g,l)}+r_{t|t-1}^{(g,l)}(1-P_\mathrm{d})).
\end{equation*}

The overall non-normalized posterior log-weight for an entire association hypothesis $\boldsymbol{\theta}_i$ is given by
\begin{equation*}
   \widetilde{l}_{t|t}^{(i)} = l_{t|t-1}^{(g)} + \sum_{j:\theta_{i,j}\neq0}\widetilde{l}_{t|t-1}^{(i,j)} + M_t \log(\lambda_c)
\end{equation*}
where 
\begin{equation*}
    \widetilde{l}_{t|t-1}^{(i,j)} = 
    \begin{cases}
        l_{t|t-1}^{(i,j)} - \log(\lambda_c) & \text{if } \theta_{i,j} = 0\\
        l_{t|t-1}^{(i,j)}                   & \text{otherwise}.
    \end{cases}
\end{equation*}
Then, the normalized log-weights can be computed as follows
\begin{equation*}
    l_{t|t}^{(i)} = \widetilde{l}_{t|t}^{(i)} - \widetilde{l}_{t|t}^{(i^*)} - \log\sum_{i}\exp\left(\widetilde{l}_{t|t}^{(i)} - \widetilde{l}_{t|t}^{(i^*)}\right)
\end{equation*}
where $i^*$ is the index related to the maximum log-weight.
%
Estimations $\hat{\mathbf{s}}_{t,n}$ are finally extracted from the posterior distribution, considering the mean value $\boldsymbol{\mu}_{t|t}^{(i,j)}$ of the \ac{MB} components with probability of existence $r_{t|t}^{(i,j)} \geq \gamma_\mathrm{e}$ from the \ac{MBM} component with highest probability $w_{t|t}^{(i)}$.


\begin{figure}[t]
    \centering    
    \includegraphics[width=1\columnwidth]{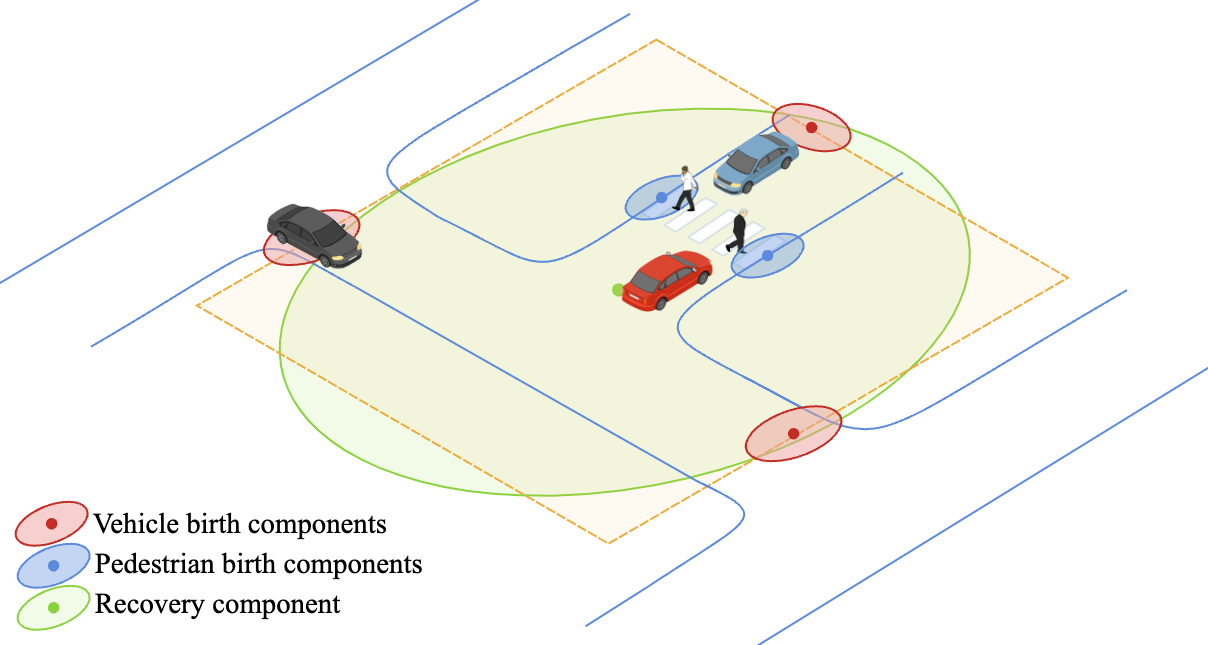}
    \caption{Birth components for tracking algorithms. Red ellipsoids stand for vehicle birth components, and blue ones represent pedestrian birth components. The green component that covers all the surveillance area is necessary to restore tracks that disappeared due to consecutive missed detections.
    }
    \label{fig:birth}
\end{figure}

\subsubsection{Birth and Death Model}
To capture new targets appearing in the surveillance area, a set of Gaussian components are generated for the \ac{PHD} filter intensity function, as well as Bernoulli components are generated for the \ac{MBM} \ac{p.d.f.}, respectively, as
\begin{align*}
    \bigl\{ w^{(b)},\boldsymbol{\mu}^{(b)},\mathbf{P}^{(b)}\bigr\}_{b=1}^{\mathcal{H}_{b}}, \,\,\,
    \bigl\{ r^{(b)},\boldsymbol{\mu}^{(b)},\mathbf{P}^{(b)}\bigr\}_{b=1}^{\mathcal{B}}.
\end{align*}
Such components can be generated considering prior information about the environment. For example, in a vehicular scenario, the number of hypotheses and their mean value, covariance, and importance are set based on the lanes and crosswalks in the environment as sketched in Fig.~\ref{fig:birth}.

Target death is modeled through a constant probability of survival $P_\mathrm{s}$. 
During prediction, if a component is associated with a missed detection, its weight is multiplied by a factor proportional to $P_\mathrm{s}$, which means that consecutive missed detections lead to unlikely target state components.

\subsubsection{Motion Model}
A linear prediction model is selected to track the behavior of both extended and point-like targets.
This is justified by the value of $T_\mathrm{scan}$ relative to the target velocity, which allows approximating the motion of the target as linear among consecutive acquisitions (see Section~\ref{sec:results}). 
The transition matrix and process noise covariance matrix adopted in \eqref{eq:kalman} and \eqref{eq:kalman2} are
$\mathbf{F} = [1, 0, T_{\mathrm{scan}},0;\,0, 1, 0, T_{\mathrm{scan}};\, 0, 0, 1, 0;\, 0, 0, 0, 1]$ and
$\mathbf{Q} = \alpha_\mathrm{q}\,T_{\mathrm{scan}}\mathbf{I}_4$, where $\alpha_\mathrm{q}$ is a parameter that represents the prediction uncertainty about the target motion.

\subsubsection{Measurement Model}
As stated, only the position information about the targets is extracted from measurements, while velocity is inferred from consecutive position measurements. Therefore, considering a linear measurement model, the  measurement matrix $  \mathbf{H} = [1, 0, 0, 0;\,
0, 1, 0, 0]$ is considered in \eqref{eq:update1} and \eqref{eq:update2}.
Due to high-resolution maps that generate more detections from each target, the measurement covariance matrix is evaluated from the detection distribution to take advantage of the information contained in multiple detections. 
Let us define the set of map points $\mathbf{L}_t$ associated to a particular measurement extracted after clustering $\mathbf{z}_{t,m}$ as $\mathbf{L}_t^{(\mathbf{z}_{t,m})}$.
The covariance measurement matrix in \eqref{eq:update1} and \eqref{eq:update2} is then evaluated as 
$\mathbf{R}_t = (N_{\mathbf{z}_{t,m}}-1)^{-1} \sum_{{n=1}}^{N_{\mathbf{z}_{t,m}}}(\mathbf{l}_{t,n}^{(\mathbf{z}_{t,m})} - \mathbf{z}_{t,m})(\mathbf{l}_{t,n}^{(\mathbf{z}_{t,m})} - \mathbf{z}_{t,m})^{{\mathsf{T}}}$,
where $N_{\mathbf{z}_{t,m}}$ represents the number of maps points associated to the measurements $\mathbf{z}_{t,m}$, and $\mathbf{l}_{t,n}^{(\mathbf{z}_{t,m})}$ is the $n$th map point in the set $\mathbf{L}_t^{(\mathbf{z}_{t,m})}$.

\subsection{Post Processing}
In order to control the complexity of the algorithms and to achieve a good estimation accuracy, a set of post processing procedures must be implemented.

In the \ac{PHD} filter, pruning, capping, and merging are sequentially implemented to reduce the number of components in the posterior intensity function.  
Firstly, pruning is implemented by removing all components in the posterior whose weights $w_{t|t}^{(h)}$ are below a predefined threshold $\gamma_\mathrm{p}$.
Then, capping is performed on the remaining components, selecting at most the $\gamma_\mathrm{q}$ components with greatest $w_{t|t}^{(h)}$ values, fixing the maximum number of components in the posterior to $\gamma_\mathrm{q}$.
Finally, on the remaining components survived after capping, merging is implemented as follows 
\begin{align*}
    w_{t|t}^{(k)} &= \sum_{i\in\boldsymbol{\zeta}_\mathrm{m}}w_{t|t}^{(i)}\,, \quad 
    \boldsymbol{\mu}_{t|t}^{(k)} = \sum_{i\in\boldsymbol{\zeta}_\mathrm{m}} w_{t|t}^{(i)}\boldsymbol{\mu}_{t|t}^{(i)} \\
    \mathbf{P}_{t|t}^{(k)} &= \sum_{i\in\boldsymbol{\zeta}_\mathrm{m}} w_{t|t}^{(i)}\mathbf{P}_{t|t}^{(i)} + (\boldsymbol{\mu}_{t|t}^{(i)}-\boldsymbol{\mu}_{t|t}^{(k)})(\boldsymbol{\mu}_{t|t}^{(i)}-\boldsymbol{\mu}_{t|t}^{(k)})^\mathsf{T}
\end{align*}
where $k$ is the new index assigned to the derived component and $i$ is the index of the merged component.
$\zeta_\mathrm{m}$ is a set containing the components that must be merged, i.e., whose mean value distance satisfies $\left\|\boldsymbol{\mu}_{t|t}^{(i)}-\boldsymbol{\mu}_{t|t}^{(j)}\right\|_2<\gamma_{\mathrm{v}}$.

In the \ac{MBM} filter, during the update phase, a gate for eligible data associations is implemented, pruning all the association hypotheses with $w_{t|t}^{(k)} < \xi_\mathrm{a}$. 
Both the \ac{MBM} and the \ac{MB} are pruned with threshold $\gamma_\mathrm{g}$ and $\gamma_\mathrm{l}$, respectively.
The residual \ac{MBM} components are capped with threshold $\gamma_\mathrm{c}$.
For the sake of increasing estimation accuracy, in the most likely \ac{MBM} component, the \ac{MB} components closer than $\gamma_\mathrm{m}$ are merged as previously described.

\begin{table}[t]
\caption{System parameters}
\begin{center}
\renewcommand{\arraystretch}{1.1}
\vspace{-3mm}
\begin{tabular}{clc} 
     \toprule
     \multicolumn{3}{c}{\textbf{Base station}}\\
     \midrule
     $\Delta_\mathrm{x}, \Delta_\mathrm{y}$ & Resampling grid resolution & $0.1$ m\\ 
     $N_\mathrm{s}$ & Number of base stations & $6$\\ 
     $T_\mathrm{scan}$ & Scan period & $0.05$ s\\ 
     $\rho_\mathrm{p}$ & Fraction of power reserved for sensing & \\ 
     $\rho_\mathrm{t}$ & Fraction of time reserved for sensing & \\
     $\rho_\mathrm{f}$ & Fraction of frequency reserved for sensing & \\
     $\gamma_\mathrm{s}$ & Sharing threshold & V$^2$/Hz\\
     $N_\mathrm{b}$ & Number of bits & 16\\
     \multicolumn{3}{l}{\vspace{-0.2cm}}\\
     \toprule
     \multicolumn{3}{c}{\textbf{Clustering}}\\
     \midrule
     $\gamma_\mathrm{d}$ & Detection threshold & $2\cdot 10^{-7}\,$V$^2$/Hz\\ 
     $\xi_\mathrm{NN}$ & Gate \ac{k-NN} & $5$ m\\ 
     $\xi_\mathrm{d}$ & Gate \ac{DBSCAN} & $3$ m\\ 
     $N_\mathrm{d}$ & Minimum number of points \ac{DBSCAN} & 50\\ 
     \multicolumn{3}{l}{\vspace{-0.2cm}}\\
     \toprule
     \multicolumn{3}{c}{\textbf{Tracking}}\\
     \midrule
     $\lambda$ & Clutter intensity & $0.001$\\ 
     $\mathbf{H}$ & Measurement matrix &\\ 
     $\mathbf{F}$ & Transition matrix &\\ 
     $\mathbf{R}_t$ & Measurements covariance &\\ 
     $\mathbf{Q}$ & Process noise covariance &\\ 
     $\alpha_\mathrm{q}$  & Prediction uncertainty & $5$\\
     $P_\mathrm{d}$ & Probability of detection & $0.99$\\ 
     $P_\mathrm{s}$ & Probability of survival & $0.9$\\
     $\gamma_\mathrm{e}$ & Existing threshold & $0.99$\\ 
     $\gamma_\mathrm{p}$ & Pruning \ac{PHD} components & $100 \cdot 10^{-6}$\\ 
     $\gamma_\mathrm{q}$ & Capping \ac{PHD} components & $10$\\ 
     $\gamma_\mathrm{v}$ & Merging \ac{PHD} components & $5$\\ 
     $\xi_\mathrm{a}$ & Gate admissible associations & $14$\\ 
     $\gamma_\mathrm{g}$ & Pruning \ac{MBM} components & $10^{-15}$\\ 
     $\gamma_\mathrm{l}$ & Pruning \ac{MB} components & $100 \cdot 10^{-6}$\\ 
     $\gamma_\mathrm{c}$ & Capping \ac{MBM} components & $10$\\ 
     $\gamma_\mathrm{m}$ & Merging \ac{MB} components & $5$\\ 
     \multicolumn{3}{l}{\vspace{-0.2cm}}\\
     \toprule
     \multicolumn{3}{c}{\textbf{\ac{OSPA} metric}}\\
     \midrule
     $p$ & \ac{OSPA} order & $2$\\ 
     $\xi_\mathrm{g}$ & \ac{OSPA} gate & $5$\\ 
     \bottomrule
\end{tabular}
\end{center}
\label{tab:track}

\end{table}

\section{Numerical Results}\label{sec:results}
In this section, we present numerical results to explore the interplay between communication and sensing. We investigate the impact of the amount of power dedicated to sensing $\rho_\mathrm{p}$, the fraction of subcarriers $\rho_\mathrm{f}$ and the fraction of time $\rho_\mathrm{t}$. Moreover, the cost/benefits of cooperation is investigated by varying the sharing threshold $\gamma_\mathrm{s}$.

\subsection{Performance Metrics}
The dual-function system is investigated under three perspectives: communication performance, communication overhead, and sensing performance. 
The communication performance is evaluated through the sum rate in \eqref{eq:Shannon}, which measures the downlink network capacity of each \ac{BS} given a particular set of resources reserved for communication. The cost of sensing is estimated in terms of sum rate reduction and network overhead, as per \eqref{eq:overhead}, due to sharing sensing data among the \acp{BS}. Since a complex scenario with multiple point-like and extended targets and multipath propagation is considered, the \ac{OSPA} metric is used to evaluate the sensing performance of the network, also in terms of cooperation gain. This metric is widely used to evaluate the performance of tracking algorithms in multi-target (or multi-scatterer) scenarios, as it allows to combine localization and detection performance into a single metric, taking into account false alarms and missed detections \cite{SchVoVo:08}.
The $p$-order \ac{OSPA} metric is defined as \cite{BeaBaBa:17,RahGarSve:17}
\begin{equation}\label{eq:OSPAdef}
        \mathrm{OSPA}\! =\!
        \Bigg[\frac{1}{N_\mathrm{c}}\Bigg(\sum_{(i,j)\in\boldsymbol{\zeta}_\mathrm{g}^*}\hspace{-8pt}\|\mathbf{\underline{s}}_{t,i} - \hat{\mathbf{\underline{s}}}_{t,j}\|_p^p+\frac{\xi_\mathrm{g}^p}{2}(|\mathbf{\underline{S}}_t|+ |\hat{\mathbf{\underline{S}}}_t|-2|\boldsymbol{\zeta}_\mathrm{g}^*|)\Bigg)\Bigg]^{\!\!\frac{1}{p}}
\end{equation}
where $\hat{\mathbf{\underline{S}}}_t$ represents the first two rows of $\hat{\mathbf{{S}}}_t$ which is a matrix containing the estimated state vectors of all the detected targets at time $t$; hence  $\hat{\mathbf{{S}}}_t$ retains only the estimated target positions. Moreover, the parameter $\xi_\mathrm{g}$ represents the \ac{OSPA} gate: estimated positions that are further than $\xi_\mathrm{g}$ from the actual target positions are considered as false alarms and, dually, real target positions that are not associated with any estimates (because not inside the gate) will be counted as missed detections.
The vector $\boldsymbol{\zeta}_\mathrm{g}^*$ represents the best assignment between the estimated set of objects $\hat{\mathbf{\underline{S}}}_t$ and the true ones $\mathbf{\underline{S}}_t$, while $|\mathbf{\underline{S}}_t|$, $|\hat{\mathbf{\underline{S}}}_t|$, and $|\boldsymbol{\zeta}_\mathrm{g}^*|$ are the cardinality of the considered set, namely, the number of objects in the estimated state vector, the ground truth cardinality, and the data association vector, respectively. Finally, $N_\mathrm{c}$ is the number of elements in the \ac{OSPA} metric given by $N_\mathrm{c} = |\mathbf{\underline{S}}_t|+ |\hat{\mathbf{\underline{S}}}_t|-|\boldsymbol{\zeta}_\mathrm{g}^*|$.

When $p=2$, the first term in \eqref{eq:OSPAdef} can be interpreted as the mean square error between the estimated and actual target positions. Instead, the second term can be rewritten as
\begin{equation*}
    \frac{\xi_\mathrm{g}^p}{2}(|\mathbf{\underline{S}}_t|- |\boldsymbol{\zeta}_\mathrm{g}^*|) + \frac{\xi_\mathrm{g}^p}{2}(|\hat{\mathbf{\underline{S}}}_t|- |\boldsymbol{\zeta}_\mathrm{g}^*|)
\end{equation*}
such that the first term is proportional to the probability of missed detections, while the second one is related to the probability of false alarms, which are defined as 
\begin{equation*}
P_\mathrm{D}= \frac{|\boldsymbol{\zeta}_\mathrm{g}^*|}{|\mathbf{\underline{S}}_t|}, \quad 
    P_\mathrm{FA} = \frac{|\hat{\mathbf{\underline{S}}}_t|- |\boldsymbol{\zeta}_\mathrm{g}^*|}{|\mathbf{\underline{S}}_t|}, \quad 
    P_\mathrm{MD}=1-P_\mathrm{D}.
\end{equation*}

\begin{figure}
    \centering
\includegraphics[width=0.8\columnwidth]{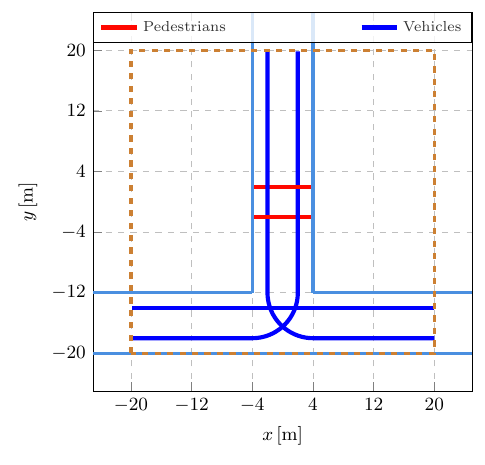}
    \caption{Targets behavior in the considered scenario. Orange dashed square represents the surveillance area while light blue lines represent lanes.}
    \label{fig:target_behavior}
\end{figure}
%
\subsection{System and Scenario Parameters}
We considered a vehicular scenario with $6$ \acp{BS} and a set of extended and point-like targets, whose behavior is depicted in Fig.~\ref{fig:target_behavior}. Pedestrians move with uniform linear motion, whereas vehicles' motion is modeled alternating static, accelerated/decelerated linear, uniform linear, and uniform circular motions. The area monitored is a typical urban crossroad of $1600\,$m$^2$, with $x \in [-20,\,20]\,$m and $y \in [-20,\,20]\,$m. The main parameters are summarized in Table~\ref{tab:track}.

The transmission parameters are: \ac{QPSK} modulation, $f_\mathrm{c} = 28\,\text{GHz}$, $\Delta f = 120\,\text{kHz}$, $K_0= 3168$ (i.e., about $400\,\text{MHz}$ bandwidth), $M = 1120$, and $M_\mathrm{s} = 112$. The \ac{EIRP} is set to $30\,$dBm, and the one-sided noise \ac{PSD} is $N_0 = 4\cdot 10^{-20}\,$W/Hz. The \acp{BS} are equipped with $N_\mathrm{T} = N_\mathrm{R} = 50$ antennas. Note that in \eqref{eq:Shannon}, without loss of generality, we assume that all \acp{UE} experience the same \ac{SNR}, and the channel is considered flat,  therefore, $\mathrm{SNR}^{(k)}_\mathrm{c}=8\,$dB for all $k$.

The multipath propagation is simulated using the 3GPP 38.901 urban micro-cell \ac{LOS} scenario \cite{3GPP_38-901}. Parameters such as $\xi^r_l$, $\phi^r_l$, $\tau^r_l$, $f^r_{\mathrm{D},l}$, $\theta^r_l$, and $\psi_l^r$ are generated using \ac{QuaDRiGa} software \cite{Lars:J14} for each target and position along the trajectories. Specifically, for each scatterer, the model considers the direct path and the $5$ strongest paths of the diffuse component, for a total of $L_\mathrm{T}=6$ paths. This results in a monostatic \ac{CIR} with $L^2_\mathrm{T}=36$ paths, as per \eqref{eq:channel-mat-OneReflectinitial}.

The six \acp{BS} are positioned along a circumference with a radius of $50\,$m, centered on the surveilled area. Their \acp{ULA} are tangent to the circumference (i.e., orthogonal to the radius), providing a scanning area of $120$° with a maximum sensing distance of $85\,$m to prevent \ac{ISI}. Each \ac{BS} performs a scan lasting $T_\mathrm{scan}=50\,$ms, and the overall scene is monitored for $10\,$s, resulting in the collection of $N_\mathrm{m} =200$ measurements (maps). The grid resolution is set to $\Delta_\mathrm{x} = 0.1\,$m and $\Delta_\mathrm{y} = 0.1\,$m. The number of bits selected to represent each map point is set to $N_\mathrm{b} = 16$. The sensing beam is periodically steered within the range $[-\Theta,\Theta]$, where $\Theta=60\,$°, with a fixed angular increment of $\Delta \Theta = 2.4$°.


In the clustering algorithm, the detection threshold is set to $\gamma_\mathrm{d} = 2\cdot10^{-7}\,$V$^2$/Hz, in \ac{k-NN} the gate is set to $\xi_\mathrm{NN} = 5$, and in \ac{DBSCAN} the cutoff distance and the minimum number of points to form the cluster are $\psi_\mathrm{d}=3$ and $N_\mathrm{d}=50$, respectively.

In both tracking algorithms, the clutter intensity is set to $\lambda_\mathrm{c} = 0.1$, the prediction uncertainty to $\alpha_\mathrm{q} = 5$, the initial components covariance to $\mathbf{P} = 0.5 \cdot \mathbf{I}_4$, and the probabilities of detection and survival are $P_\mathrm{d} = 0.99$ and $P_\mathrm{s} = 0.9$, respectively.

In the \ac{PHD} filter, the components pruning threshold is $\gamma_\mathrm{p} = 100\cdot10^{-6}$ while the maximum number of components is fixed to $\gamma_\mathrm{q} = 10$. The birth components for new appearing objects are initialized with importance $w^{(b)}=0.1$ and covariance $\mathbf{P}^{(b)} = 0.1\cdot\mathbf{I}_4$, with position $\boldsymbol{\mu}^{(b)}$ reflecting the components depicted in Fig~\ref{fig:birth}. The recovery component is initialized with importance $w^{(b)}=0.01$ and covariance $\mathbf{P}^{(b)} = 5\cdot\mathbf{I}_4$. The merging threshold is set to $\gamma_\mathrm{v} = 5$.

In the \ac{MBM} filter, the pruning threshold on the probability of existence is $\gamma_\mathrm{l} = 100\cdot 10^{-6}$ while the pruning threshold on the \ac{MBM} components is set to $\gamma_\mathrm{g} = 10^{-15}$. The maximum number of \ac{MBM} components is fixed to $\gamma_\mathrm{c} = 10$. The gate for the admissible associations is set to $\xi_\mathrm{a} = 14$. The existence threshold is set to $\gamma_\mathrm{e} = 0.99$. The birth components for new appearing objects are initialized with a probability of existence $r^{(b)}=10^{-4}$ and covariance $\mathbf{P}^{(b)} = 0.1\cdot\mathbf{I}_4$, with position $\boldsymbol{\mu}^{(b)}$ reflecting again the components reported in Fig~\ref{fig:birth}. The recovery component is initialized with importance $r^{(b)}=10^{-5}$ and covariance $\mathbf{P}^{(b)} = 5\cdot\mathbf{I}_4$. The merging threshold is set to $\gamma_\mathrm{m} = 5$.

\begin{figure}[t]
    \centering    
    \includegraphics[width=1\columnwidth]{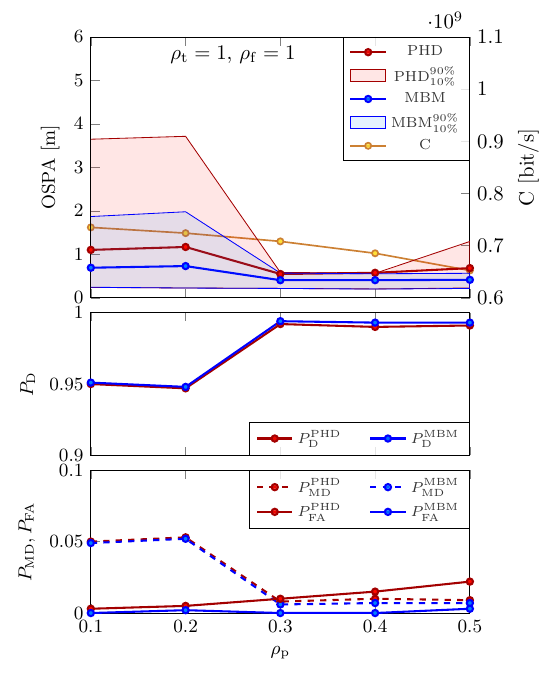}
    \caption{Communication and sensing performance varying the fraction of power reserved for \ac{JSC} $\rho_\mathrm{p}$. The downlink capacity is per \ac{BS}.
    }
    \label{fig:rhop}
\end{figure}

\begin{figure}[t]
    \centering    
    \includegraphics[width=1\columnwidth]{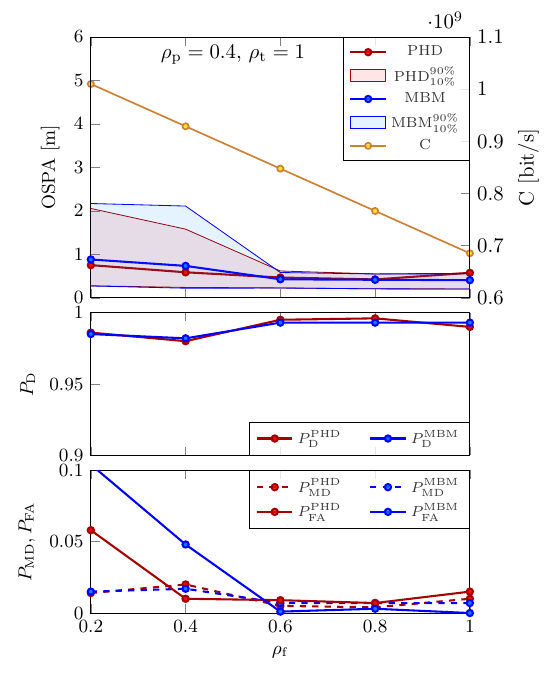}
    \caption{Communication and sensing performance varying the fraction of subcarriers reserved for \ac{JSC} $\rho_\mathrm{f}$. The downlink capacity is per \ac{BS}.
    }
    \label{fig:rhof}
\end{figure}

\begin{figure}[t]
    \centering    
    \includegraphics[width=1\columnwidth]{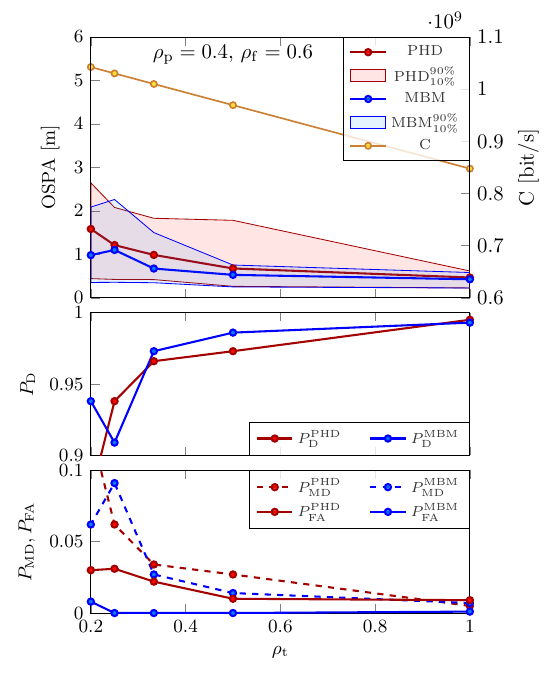}
    \caption{Communication and sensing performance varying the fraction of time reserved for \ac{JSC} $\rho_\mathrm{t}$. The downlink capacity is per \ac{BS}.
    }
    \label{fig:rhot}
\end{figure}

\begin{figure}[t]
    \centering    
    \includegraphics[width=1\columnwidth]{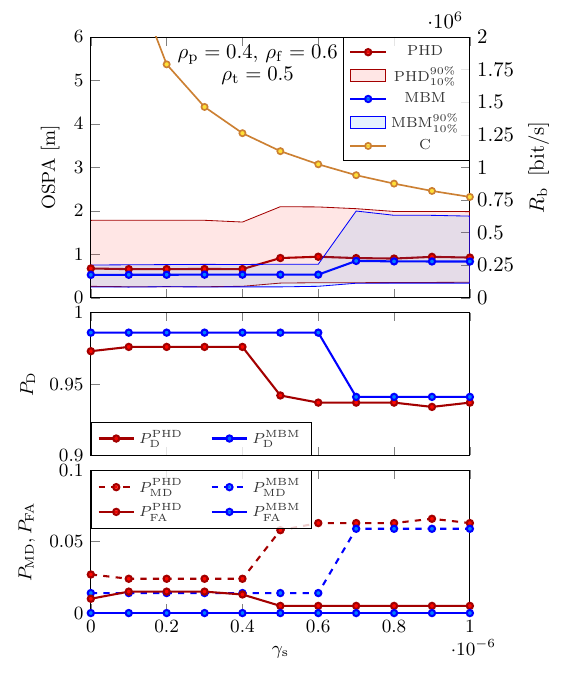}
    \caption{Network overhead due to cooperation and sensing performance varying the sharing threshold $\gamma_\mathrm{s}$. The data rate is per \ac{BS}.
    }
    \label{fig:gammas}
\end{figure}

\begin{figure}[t]
    \centering    
    \includegraphics[width=1\columnwidth]{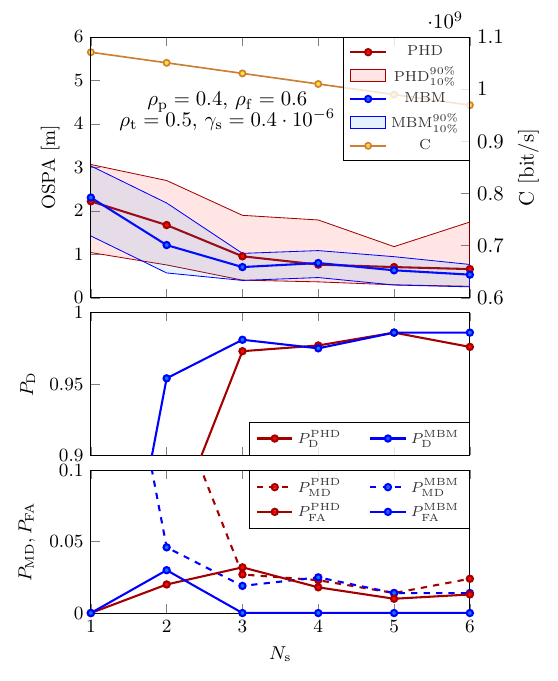}
    \caption{Communication and sensing performance varying the number of \acp{BS} adopted for sensing $N_\mathrm{s}$. The downlink capacity is per \ac{BS}.
    }
    \label{fig:sensors}
\end{figure}

\subsection{Impact of the Fraction of Power $\rho_\mathrm{p}$}
In Fig.~\ref{fig:rhop}, the tracking and communication performance are reported by varying the fraction of power dedicated to sensing, $\rho_\mathrm{p}$, when $\rho_\mathrm{f} = 1$ and $\rho_\mathrm{t} = 1$.

At the top, the average \ac{OSPA} metric is displayed for both tracking algorithms in red and blue, respectively. The yellow curve represents the downlink sum rate defined in \eqref{eq:Shannon}. The blue and red areas represent the range of values between the $90$th and $10$th percentiles of the \ac{OSPA} distance. The detection probability of the tracking algorithms is shown in the middle plot, while the missed detection probability and false alarm probability are presented at the bottom. This structure is repeated for all subsequent figures.

Both \ac{PHD} and \ac{MBM} filters perform similarly for varying $\rho_\mathrm{p}$. Specifically, at low values of $\rho_\mathrm{p}$, weak targets such as pedestrians disappear, leading to increased missed detection.
It is interesting to note that for large values of $\rho_\mathrm{p}$ (e.g., $0.5$), the localization error experiences a slight increase. This is due to increased false alarms caused by multipath propagation, which becomes more relevant by increasing the sensing transmission power.

From a communication perspective, increasing $\rho_\mathrm{p}$ results in lower downlink capacity, as expected. For a value of $\rho_\mathrm{p} = 0.1$, the localization error is around $1\,$m, while the downlink capacity is $0.72\,$Gbit/s. Increasing $\rho_\mathrm{p}$ to $0.4$ halves the \ac{OSPA} error while reducing the network capacity to $0.68\,$Gbps, which represents a $6\,\%$ loss in communication performance.

\subsection{Impact of the Fraction of Subcarriers $\rho_\mathrm{f}$}
Communication and sensing performance, varying the frequency resources allocation between the two functions, are depicted in Fig.~\ref{fig:rhof}, considering $\rho_\mathrm{p} = 0.4$ and $\rho_\mathrm{t} = 1$.

From a sensing perspective, it is interesting to observe that as $\rho_\mathrm{f}$ decreases, the \ac{OSPA} increases due to a higher probability of false alarm. This effect can be attributed to a lower map resolution due to a lower bandwidth, resulting in target smearing that can more likely generate false target spawns.

For communication, a decrease of $\rho_\mathrm{f}$ results in an increase of the fraction of subcarriers exclusively dedicated to communication, which leads to a higher downlink sum rate. A capacity of $1\,$Gbit/s is ensured by fixing $\rho_\mathrm{f} = 0.2$, with a localization error around $0.9\,$m. Increasing $\rho_\mathrm{f}$ to $0.6$, the localization error drops to $0.5\,$m, with a reduction in network capacity of $15\,$\%, i.e., $0.15\,$Gbit/s.

\subsection{Impact of the Fraction of Time $\rho_\mathrm{t}$}
In Fig.~\ref{fig:rhot}, communication and sensing performance are presented, varying the fraction of time devoted to \ac{JSC}, $\rho_\mathrm{t}$, with $\rho_\mathrm{p} = 0.4$ and $\rho_\mathrm{f} = 0.6$.

Notably, as $\rho_\mathrm{t}$ decreases, an increase in the \ac{OSPA} error for both tracking algorithms can be observed. This increase can be attributed to a higher probability of missed detections, due to less accurate motion prediction. In this setup, the \ac{MBM} tracking algorithm demonstrates more accurate and stable performance compared to the \ac{PHD} filter. This advantage arises from the \ac{MBM} tracking algorithm's enhanced capability to handle missing measurements due to less frequent map acquisition.

Additionally, as the fraction of time allocated for \ac{JSC}, $\rho_\mathrm{t}$, increases, there is a decrease in the downlink sum rate. This decrease results from a lower fraction of time when all communication resources are available, limiting the overall downlink capacity. With $\rho_\mathrm{t}$ fixed at $0.5$, the \ac{OSPA} metric is around $0.5\,$m, while the downlink capacity is $0.97\,$Gbit/s. Halving $\rho_\mathrm{t}$, the capacity increases to $1.03\,$Gbit/s, with a gain of $6\,$\%, while the localization performance decreases, with an \ac{OSPA} distance around $1.2\,$m.

\subsection{Sharing Threshold $\gamma_\mathrm{s}$ and Network Overhead}
To investigate the impact of cooperation on both communication and sensing functionalities, in this section, we analyze their performance by varying the sharing threshold $\gamma_\mathrm{s}$. The resources allocated for sensing are $\rho_\mathrm{p} = 0.4$, $\rho_\mathrm{t} = 0.6$, and $\rho_\mathrm{t} = 0.5$. Remarkably, cooperative sensing with this resource configuration enables achieving localization errors below $0.8\,$m and a \ac{BS} downlink capacity exceeding $1\,$Gbit/s.

As shown in Fig.~\ref{fig:gammas}, increasing the sharing threshold $\gamma_\mathrm{s}$ leads to a rapid decrease in the amount of data that \acp{BS} need to transfer to the \ac{FC} compared to sharing the entire maps (i.e., $\gamma_\mathrm{s}=0$). Notably, when $\gamma_\mathrm{s}<0.1\cdot 10^{-6}$ V$^2$/Hz, the data rate increases abruptly because most of the noise samples in the range-angle maps are shared with the \ac{FC}. By choosing an appropriate sharing threshold, a significant reduction in overhead is achieved, resulting in a required data rate $R_\mathrm{b} = 613\,$Kbit/s to share the selected map points compared to $R_\mathrm{b} = 120\,$Mbit/s when sharing the entire maps.

From the sensing perspective, increasing $\gamma_\mathrm{s}$ maintains stable system performance until the number of points filtered by the threshold becomes excessively high. In such situations, weak targets may go undetected, leading to an increase in the \ac{OSPA} metric.
Furthermore, in this setup, the \ac{MBM} filter shows greater robustness to the lack of data caused by thresholding compared to \ac{PHD} tracking. This characteristic allows for setting a larger threshold $\gamma_\mathrm{s}$ with benefits in terms of cooperation overhead.

\subsection{Impact of Number of Sensors $N_\mathrm{s}$}
Communication and sensing performance, varying the number of \acp{BS} dedicated to sensing, are presented in Fig.~\ref{fig:sensors}, considering $\rho_\mathrm{p} = 0.4$, $\rho_\mathrm{f} = 0.6$, $\rho_\mathrm{t} = 0.5$, and $\gamma_\mathrm{s} = 0.4\cdot10^{-6}$ V$^2$/Hz.

From a sensing perspective, it is important to note the increase in the \ac{OSPA} error as the number of sensors $N_\mathrm{s}$ decreases. Moreover, the minimum number of sensors required to guarantee a localization error lower than $1\,$m is $3$, and for a lower number of sensors, the error rapidly increases along with the detection probability. \ac{PHD} and \ac{MBM} exhibit  comparable performance varying the number of sensors.

From a communication point of view, decreasing $N_\mathrm{s}$ slightly increases the downlink capacity. This increase is due to an increment of the communication resources that arise from \acp{BS} not selected to perform sensing.

\section{Conclusion}\label{sec:conclusion}
In this work, we presented a \ac{JSC} framework for \ac{OFDM} systems, leveraging the infrastructure of mobile radio networks to enable cooperative sensing among \acp{BS}. We proposed a strategy to fuse soft maps acquired from \acp{BS} at a \ac{FC}, along with a thresholding method for managing the amount of data shared between \acp{BS} and the \ac{FC}. Two tracking algorithms, \ac{PHD} and \ac{MBM} filters, were employed, combined with a tailor-made three-step clustering strategy to address point-like and extended targets, specifically pedestrians and cars.

The overall networked system was tested by varying the fraction of resources allocated for sensing, including power ($\rho_\mathrm{p}$), frequency ($\rho_\mathrm{f}$), and time ($\rho_\mathrm{t}$), while also considering network overhead managed through map thresholding.
Localization performance was assessed through the \ac{OSPA} metric, as well as the probability of detection and false alarm, while communication performance was evaluated through \ac{BS} downlink sum rate.

Numerical results have shown that through the cooperation between \acp{BS}, the localization of both extended and point-like targets is possible with an error of less than $80\,cm$, while ensuring a downlink capacity greater than $1\,$Gbit/s, considering $\rho_\mathrm{p}=0.4$, $\rho_\mathrm{f}=0.6$, and $\rho_\mathrm{t}=0.5$.
Through map management, the aforementioned results can be obtained with a network overhead of $613\,$kbit/s, reducing the amount of data shared among \acp{BS} by a factor of $190$ compared to $120\,$Mbit/s without any map filtering.

\balance
\bibliographystyle{IEEEtran}
\bibliography{IEEEabrv,bibliography}

\begin{thebibliography}{10}
\providecommand{\url}[1]{#1}
\csname url@samestyle\endcsname
\providecommand{\newblock}{\relax}
\providecommand{\bibinfo}[2]{#2}
\providecommand{\BIBentrySTDinterwordspacing}{\spaceskip=0pt\relax}
\providecommand{\BIBentryALTinterwordstretchfactor}{4}
\providecommand{\BIBentryALTinterwordspacing}{\spaceskip=\fontdimen2\font plus
\BIBentryALTinterwordstretchfactor\fontdimen3\font minus
  \fontdimen4\font\relax}
\providecommand{\BIBforeignlanguage}[2]{{%
\expandafter\ifx\csname l@#1\endcsname\relax
\typeout{** WARNING: IEEEtran.bst: No hyphenation pattern has been}%
\typeout{** loaded for the language `#1'. Using the pattern for}%
\typeout{** the default language instead.}%
\else
\language=\csname l@#1\endcsname
\fi
#2}}
\providecommand{\BIBdecl}{\relax}
\BIBdecl

\bibitem{BarLiuWinCon:J22}
S.~Bartoletti, Z.~Liu, M.~Z. Win, and A.~Conti, ``Device-free localization of
  multiple targets in cluttered environments,'' \emph{{IEEE} Trans. Aerosp.
  Electron. Syst.}, vol.~58, no.~5, pp. 3906--3923, Oct. 2022.

\bibitem{Tho:C21}
R.~Thomä, T.~Dallmann, S.~Jovanoska, P.~Knott, and A.~Schmeink, ``Joint
  communication and radar sensing: An overview,'' in \emph{Proc. Europ. Conf.
  on Ant. and Prop. (EuCAP)}, Dusseldorf, Germany, Mar. 2021, pp. 1--5.

\bibitem{Sch:C20}
S.~Schieler, C.~Schneider, C.~Andrich, M.~Döbereiner, J.~Luo, A.~Schwind,
  R.~S. Thomä, and G.~Del~Galdo, ``{OFDM} waveform for distributed radar
  sensing in automotive scenarios,'' \emph{Int. J. of Microw. and Wireless
  Tech.}, vol.~12, no.~8, p. 716–722, 2020.

\bibitem{KwoLiuConParWin:J23}
G.~Kwon, Z.~Liu, A.~Conti, H.~Park, and M.~Z. Win, ``Integrated localization
  and communication for efficient millimeter wave networks,'' \emph{{IEEE} J.
  Sel. Areas Commun.}, vol.~41, no.~12, Dec. 2023.

\bibitem{MorRazWinCon:J23}
F.~Morselli, S.~M. Razavi, M.~Z. Win, and A.~Conti, ``Soft information based
  localization for {5G} networks and beyond,'' \emph{{IEEE} Trans. Wireless
  Commun.}, vol.~22, pp. 1--16, 2023.

\bibitem{FenWeiChe:21}
Z.~Feng, Z.~Wei, X.~Chen, H.~Yang, Q.~Zhang, and P.~Zhang, ``Joint
  communication, sensing, and computation enabled {6G} intelligent machine
  system,'' \emph{{IEEE} Netw.}, vol.~35, no.~6, pp. 34--42, 2021.

\bibitem{JohVenGroLop:J22}
J.~Johnston, L.~Venturino, E.~Grossi, M.~Lops, and X.~Wang, ``{MIMO OFDM}
  dual-function radar-communication under error rate and beampattern
  constraints,'' vol.~40, no.~6, pp. 1951--1964, 2022.

\bibitem{Cui:21IntegratingSA}
Y.~Cui, F.~Liu, X.~Jing, and J.~Mu, ``Integrating sensing and communications
  for ubiquitous {IoT}: Applications, trends, and challenges,'' \emph{{IEEE}
  Netw.}, vol.~35, pp. 158--167, 2021.

\bibitem{LiuZhaoWang:19}
F.~Liu, P.~Zhao, and Z.~Wang, ``{EKF}-based beam tracking for {mmWave} {MIMO}
  systems,'' \emph{{IEEE} Commun. Lett.}, vol.~23, no.~12, pp. 2390--2393,
  2019.

\bibitem{ZhenLiu:22}
Z.~Du, F.~Liu, and Z.~Zhang, ``Sensing-assisted beam tracking in {V2I}
  networks: Extended target case,'' in \emph{Proc. IEEE Int. Conf. on
  Acoustics, Speech and Signal Process. (ICASSP)}, May 2022, pp. 8727--8731.

\bibitem{Favarelli:C23}
E.~Favarelli, E.~Matricardi, L.~Pucci, E.~Paolini, W.~Xu, and A.~Giorgetti,
  ``Map fusion and heterogeneous objects tracking in joint sensing and
  communication networks,'' in \emph{Proc. European Radar Conf. (EuRAD)},
  Berlin, Germany, Sep. 2023.

\bibitem{Favarelli:D23}
------, ``Sensor fusion and extended multi-target tracking in joint sensing and
  communication networks,'' in \emph{Proc. IEEE Int. Conf. on Commun. (ICC)},
  Rome, Italy, May 2023.

\bibitem{Favarelli:E23}
------, ``Multi-base station cooperative sensing with ai-aided tracking,'' in
  \emph{arXiv}, Oct. 2023.

\bibitem{PucPaoGio:J22}
L.~{Pucci}, E.~{Paolini}, and A.~{Giorgetti}, ``System-level analysis of joint
  sensing and communication based on {5G} new radio,'' in \emph{{IEEE} J. Sel.
  Areas Commun.}, vol.~40, no.~7, July 2022, pp. 2043--2055.

\bibitem{asplund2020advanced}
H.~Asplund, D.~Astely, P.~von Butovitsch, T.~Chapman, M.~Frenne,
  F.~Ghasemzadeh, M.~Hagstr{\"o}m, B.~Hogan, G.~J{\"o}ngren, J.~Karlsson
  \emph{et~al.}, \emph{Advanced Antenna Systems for 5G Network Deployments:
  Bridging the Gap Between Theory and Practice}.\hskip 1em plus 0.5em minus
  0.4em\relax Academic Press, 2020.

\bibitem{FavMatPuc:22}
E.~Favarelli, E.~Matricardi, L.~Pucci, E.~Paolini, W.~Xu, and A.~Giorgetti,
  ``Tracking and data fusion in joint sensing and communication networks,'' in
  \emph{Proc. IEEE Globecom Workshops}, Rio de Janeiro, Brasil, Dec. 2022, pp.
  341--346.

\bibitem{richards}
M.~A. Richards, \emph{Fundamentals of radar signal processing}.\hskip 1em plus
  0.5em minus 0.4em\relax McGraw-Hill, 2005.

\bibitem{Fur:J21a}
M.~F. Keskin, V.~Koivunen, and H.~Wymeersch, ``Limited feedforward waveform
  design for {OFDM} dual-functional radar-communications,'' \emph{{IEEE} Trans.
  Signal Process.}, vol.~69, pp. 2955--2970, 2021.

\bibitem{Fur:J21b}
M.~F. Keskin, H.~Wymeersch, and V.~Koivunen, ``{MIMO-OFDM} joint
  radar-communications: Is ici friend or foe?'' \emph{{IEEE} J. Sel. Topics in
  Signal Process.}, vol.~15, no.~6, pp. 1393--1408, 2021.

\bibitem{alhassoun2019theoretical}
M.~Alhassoun and G.~D. Durgin, ``A theoretical channel model for spatial fading
  in retrodirective backscatter channels,'' \emph{{IEEE} Trans. Wireless
  Commun.}, vol.~18, no.~12, pp. 5845--5854, 2019.

\bibitem{blunt2011performance}
S.~D. Blunt, J.~G. Metcalf, C.~R. Biggs, and E.~Perrins, ``Performance
  characteristics and metrics for intra-pulse radar-embedded communication,''
  \emph{{IEEE} J. Sel. Areas Commun.}, vol.~29, no.~10, pp. 2057--2066, 2011.

\bibitem{Sko:B08}
M.~I. Skolnik, \emph{Radar handbook}.\hskip 1em plus 0.5em minus 0.4em\relax
  McGraw-Hill Education, 2008.

\bibitem{buhren:06Automotive}
M.~B{\"u}hren and B.~Yang, ``Automotive radar target list simulation based on
  reflection center representation of objects,'' in \emph{Proc. Int. Work. on
  Intelligent Transp. (WIT)}, Hamburg, Germany, Mar. 2006, pp. 161--166.

\bibitem{BuhrenYang:06}
------, ``Simulation of automotive radar target lists using a novel approach of
  object representation,'' in \emph{Proc. IEEE Intelligent Veh. Symp.},
  Meguro-Ku, Japan, 2006, pp. 314--319.

\bibitem{Braun}
M.~Braun, ``{OFDM} radar algorithms in mobile communication networks,'' Ph.D.
  dissertation, Karlsruhe Institute of Technology, 2014.

\bibitem{RodBlaColLom:J23}
J.~T. Rodriguez, G.~P. Blasone, F.~Colone, and P.~Lombardo, ``Exploiting the
  properties of reciprocal filter in low-complexity {OFDM} radar signal
  processing architectures,'' \emph{{IEEE} Trans. Aerosp. Electron. Syst.},
  vol.~59, no.~5, pp. 6907--6922, 2023.

\bibitem{FullDuplex}
C.~B. Barneto, T.~Riihonen, M.~Turunen, L.~Anttila, M.~Fleischer, K.~Stadius,
  J.~Ryyn{\"a}nen, and M.~Valkama, ``Full-duplex {OFDM} radar with {LTE} and
  {5G NR} waveforms: challenges, solutions, and measurements,'' \emph{{IEEE}
  Trans. Microw. Theory Tech.}, vol.~67, no.~10, pp. 4042--4054, 2019.

\bibitem{HenManArnBri:C22}
M.~Henninger, S.~Mandelli, M.~Arnold, and S.~Ten~Brink, ``A computationally
  efficient {2D MUSIC} approach for {5G} and {6G} sensing networks,'' in
  \emph{Proc. IEEE Wireless Comm. and Netw. Conf. (WCNC)}, Austin, TX, USA,
  Apr. 2022, pp. 210--215.

\bibitem{XioLiu:J23}
Y.~Xiong, F.~Liu, Y.~Cui, W.~Yuan, T.~X. Han, and G.~Caire, ``On the
  fundamental tradeoff of integrated sensing and communications under
  {G}aussian channels,'' \emph{{IEEE} Trans. Inf. Theory}, vol.~69, no.~9, pp.
  5723--5751, 2023.

\bibitem{ChoSheHon:19}
H.~Y. Chongjun~Ouyang, Sheng~Wu, ``Mutual information approximation,''
  \emph{arXiv}, 2019.

\bibitem{WatBorKat:16}
J.~Watt, R.~Borhani, and A.~K. Katsaggelos, \emph{Machine Learning
  Refined}.\hskip 1em plus 0.5em minus 0.4em\relax Cambridge University Press,
  2016.

\bibitem{EstKriXia:96}
M.~Ester, H.-P. Kriegel, J.~Sander, and X.~Xu, ``A density-based algorithm for
  discovering clusters in large spatial databases with noise,'' in \emph{Proc.
  Int. Conf. on Know. Disc. in Data Mining}, Portland, Oregon, Aug. 1996, pp.
  226--231.

\bibitem{Mah:03}
R.~Mahler, ``Multitarget {Bayes} filtering via first-order multitarget
  moments,'' \emph{{IEEE} Trans. Aerosp. Electron. Syst.}, vol.~39, no.~4, pp.
  1152--1178, 2003.

\bibitem{VoMa:06}
B.-N. Vo and W.-K. Ma, ``The gaussian mixture probability hypothesis density
  filter,'' \emph{{IEEE} Trans. Signal Process.}, vol.~54, no.~11, pp.
  4091--4104, 2006.

\bibitem{GarXiaGra:19}
A.~F. García-Fernández, Y.~Xia, K.~Granström, L.~Svensson, and J.~L.
  Williams, ``Gaussian implementation of the multi-bernoulli mixture filter,''
  in \emph{Proc. Int. Conf. on Inf. Fusion (FUSION)}, Ottawa, Canada, 2019, pp.
  1--8.

\bibitem{GarWilGra:18}
A.~F. García-Fernández, J.~L. Williams, K.~Granström, and L.~Svensson,
  ``Poisson multi-bernoulli mixture filter: Direct derivation and
  implementation,'' \emph{{IEEE} Trans. Aerosp. Electron. Syst.}, vol.~54,
  no.~4, pp. 1883--1901, 2018.

\bibitem{ThaAleRub:21}
L.~B. Thanh, P.~D. Alexandrovich, and P.~Ruben, ``Multi-object multi-sensor
  tracking simulation using poisson multi-bernoulli mixture filter,'' in
  \emph{Proc. Int. Conf. on Digital Signal Process. and its Applications
  (DSPA)}, Moscow, Russian Federation, 2021, pp. 1--6.

\bibitem{SchVoVo:08}
D.~Schuhmacher, B.-T. Vo, and B.-N. Vo, ``A consistent metric for performance
  evaluation of multi-object filters,'' \emph{IEEE Trans. Signal Process.},
  vol.~56, no.~8, pp. 3447--3457, Aug. 2008.

\bibitem{BeaBaBa:17}
M.~Beard, B.~T. Vo, and B.-N. Vo, ``{OSPA(2)}: Using the {OSPA} metric to
  evaluate multi-target tracking performance,'' in \emph{Proc. Int. Conf. on
  Control, Autom. and Inf. Sci. (ICCAIS)}, Chiang Mai, Thailand, 2017, pp.
  86--91.

\bibitem{RahGarSve:17}
A.~S. Rahmathullah, A.~F. García-Fernández, and L.~Svensson, ``Generalized
  optimal sub-pattern assignment metric,'' in \emph{Proc. Int. Conf. on Inf.
  Fusion (FUSION)}, Xi'an, China, 2017, pp. 1--8.

\bibitem{3GPP_38-901}
\emph{Study on channel model for frequencies from 0.5 to 100 GHz}, 3GPP TR
  38.901, 2019, version 16.1.0.

\bibitem{Lars:J14}
S.~Jaeckel, L.~Raschkowski, K.~Börner, and L.~Thiele, ``Quadriga: A 3-d
  multi-cell channel model with time evolution for enabling virtual field
  trials,'' \emph{{IEEE} Trans. Antennas Propag.}, vol.~62, no.~6, pp.
  3242--3256, 2014.

\end{thebibliography}

\end{document}